\documentclass[12pt]{article}

\usepackage{newtxtext,newtxmath}
\usepackage{booktabs}

\usepackage{graphicx}

\usepackage[letterpaper,margin=1in]{geometry}

\renewenvironment{abstract}
	{\quotation}
	{\endquotation}

\date{}

\makeatletter
\renewcommand{\fnum@figure}{\textbf{Figure \thefigure}}
\renewcommand{\fnum@table}{\textbf{Table \thetable}}
\makeatother

\usepackage{scicite}

\usepackage{url}

\def\scititle{
	Magnon Field-Effect Transistor
}
\title{\bfseries \boldmath \scititle}

\author{
	Qian Song$^{1,2,3\ast}$,
    Sergei Prokhorenko$^{4}$,
    Thomas Cleveland$^{4}$,
    Ashish Omar$^{1}$,\and
    Zhi Yao$^{5}$,
    Sajid Husain$^{1,2}$,
    Peter Meisenheimer$^{1}$,
    Bowen Hao$^{6,7}$,\and
    Ziye Zhu$^{6,7}$,
    Pushpendra Gupta$^{1}$,
    Shashank Ojha$^{3,8}$,
    Nicholas Reiterer$^{1}$,\and
    Maya Ramesh$^{9}$,
    Sergio Puebla$^{1}$,
    Darrell Schlom$^{9}$,
    Tong Zhou$^{5,6}$,\and
    Laurent Bellaiche$^{4,10}$,
    Lane W. Martin$^{3,8,11}$,
	Ramamoorthy Ramesh$^{1,2\ast}$\and
    \small$^{1}$Department of Material Science and Engineering, University of California, Berkeley, CA, USA.\and\
	\small$^{2}$Department of Physics, University of California, Berkeley, CA, USA.\and
	\small$^{3}$Department of Material Science and NanoEngineering, Rice University, Houston, TX, USA.\and
    \small$^{4}$Smart Ferroic Materials Center, Department of Physics and Arkansas Materials Institute, \and
    \small University of Arkansas, Fayetteville, AR, USA.\and
    \small$^{5}$School of Electrical and Computer Engineering, Georgia Institute of Technology, Atlanta, GA, USA.\and
    \small$^{6}$Ningbo Institute of Digital Twin, Eastern Institute of Technology, Ningbo, Zhejiang, China\and
    \small$^{7}$Department of Physics, University of Science and Technology of China, Hefei, Anhui, China\and
    \small$^{8}$Rice Advanced Materials Institute, Rice University, Houston, TX, USA.\and
    \small$^{9}$Department of Material Science and Engineering, Cornell University, Ithaca, NY, USA.\and
    \small$^{10}$Department of Materials Science and Engineering, Tel Aviv University, Ramat Aviv, Tel Aviv, Israel\and
    \small$^{11}$Departments of Chemistry and Physics and Astronomy, Rice University, Houston, TX, USA.\and
	\small$^\ast$Corresponding author. Email: qiansong@berkeley.edu, rramesh@berkeley.edu
}

\begin{document} 

\maketitle
\newpage
\begin{abstract} \bfseries \boldmath
Abstract: Antiferromagnetic magnons offer a promising platform for high-speed, energy-efficient information processing. Intrinsic odd-parity spin splitting in a cycloidal magnetic order offers a route to nonvanishing magnon transport in compensated magnets. Here, we propose a magnonic field-effect device (akin to a transistor) based on a ferroelectric antiferromagnet BiFeO$_3$ in which a gate electric field reconfigures domain boundaries to amplify/modulate magnon transport. These boundaries switch the nonlocal inverse-spin-Hall response between low- and high-signal states, with a pronounced dependence on the direction of magnon propagation. Source-drain reversal, magnetic-field measurements, symmetry analysis, and first-principles calculations reveal nonreciprocal magnon transport associated with the reversal of spin angular momentum across the domain boundaries. Our results establish an electrically programmable, non-volatile magnon transistor. 
\end{abstract}

\noindent
\paragraph{Introduction.}
Field-effect transistors are the fundamental building blocks of modern electronics, underpinning processors, memory, and digital-logic circuits. Increasing device density and operating speeds, however, pose growing challenges for power dissipation and thermal management, limiting further device scaling. Magnons, the collective excitations of magnetically ordered spins, carry angular momentum and phase without transporting electric charge, offering a promising platform for high-speed, energy-efficient information processing\cite{chumak_magnon_2015, kajiwara_transmission_2010, barman_magnonics_2021, cornelissen_longdistance_2015}. Unlike volatile charge transport in semiconductors, which can be readily controlled by electrostatic gating, electrical manipulation of magnon transport remains challenging owing to the weak coupling between electric fields and magnetic moments in conventional magnetic materials \cite{chumak_transistor_2014, rana_magnonic_2019, cheng_nonvolatile_2024}. Multiferroics, in which ferroelectric and magnetic orders coexist and are strongly coupled, provide an alternative route to electrical-field control of antiferromagnetism \cite{eerenstein_multiferroic_2006, ramesh_multiferroics_2007, cheong_multiferroics_2007, matsukura_control_2015, spaldin_advances_2019}. In multiferroic BiFeO$_3$ (BFO), the ferroelectric polarization is coupled to the cycloidal propagation vector and magnetic chirality, enabling electrical control of magnon modes \cite{rovillain_electric_2010, bhattacharjee_ultrafast_2014, gross_realspace_2017, meisenheimer_switching_2024}. Recent studies have demonstrated electrically switchable magnon transport in BFO-based devices\cite{huang_manipulating_2024, husain_nonvolatile_2024, harris_symmetry_2025, husain_magnon_2026}. Here, we propose a field-effect magnon transistor in which an independent gate electrode creates and reconfigures antiferromagnetic domain boundaries, enabling gate-controlled enhancement of nonvolatile magnon transport through a nonreciprocal mechanism.

\begin{figure}
	\centering
	\includegraphics[width=\textwidth]{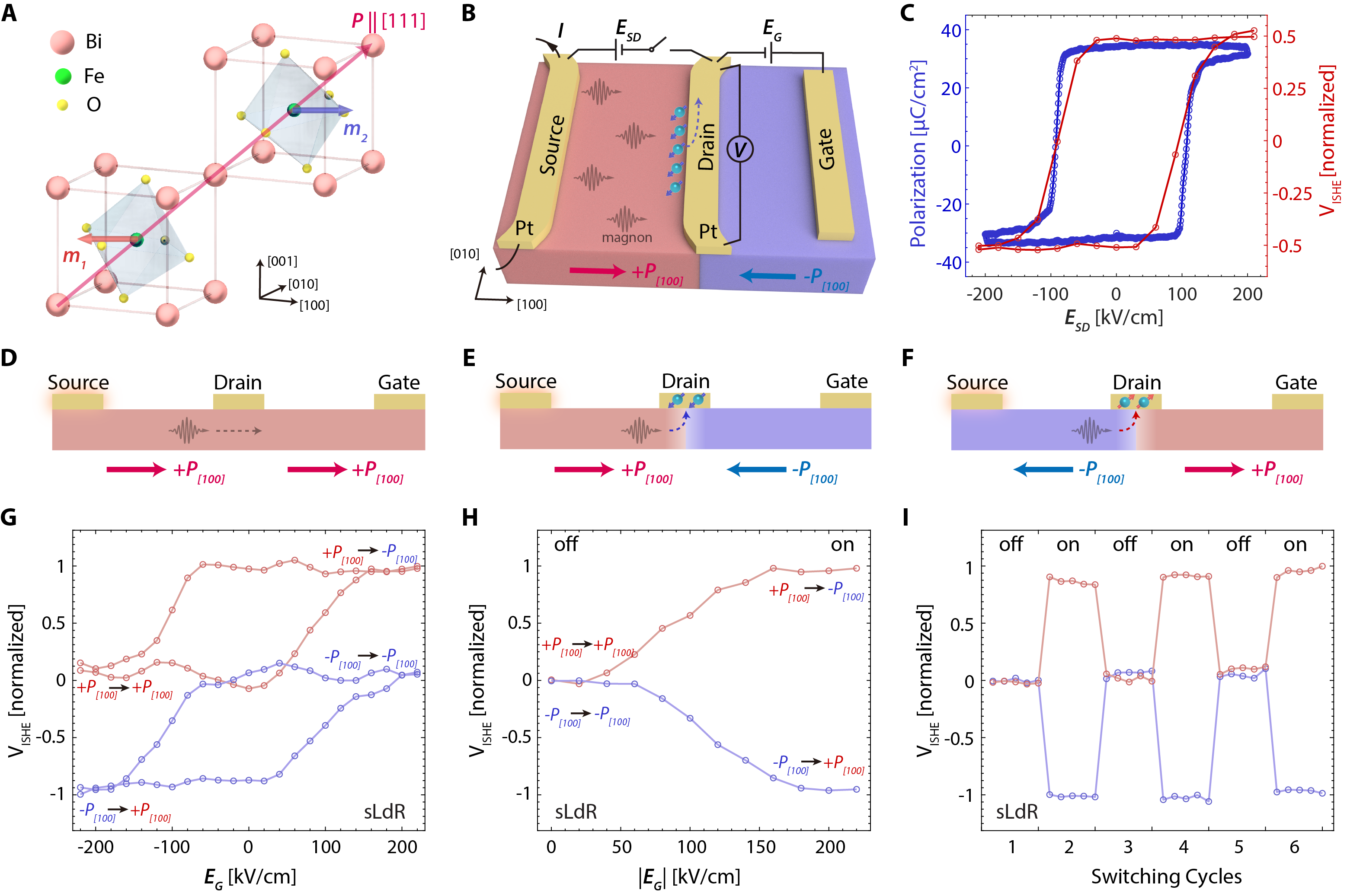}
	\caption{\textbf{Gate-controlled magnon transport in multiferroic BiFeO$_3$.}
    \textbf{A}, Crystal and magnetic structures of BiFeO$_3$, with ferroelectric polarization $P$ along [111] and antiferromagnetic sublattice moments $m_1$ and $m_2$.
    \textbf{B}, Device schematic. Magnons generated by the spin Seebeck effect propagate from source to drain and are detected via the inverse-spin-Hall effect (ISHE) in platinum (Pt) electrode. $E_{\mathrm{SD}}$ and $E_{\mathrm{G}}$ denote the electric fields applied between the source and drain and between the gate and drain, respectively. $+P_{[100]}$ (red) and $-P_{[100]}$ (blue) denote the projected polarization along [100].
    \textbf{C}, Ferroelectric polarization (blue) and ISHE (red) hysteresis loops.
    \textbf{D-F}, Magnon propagation from left to right through a single antiferromagnetic domain (\textbf{D}) and across $+P_{[100]} \rightarrow -P_{[100]}$ (\textbf{E}) and $-P_{[100]} \rightarrow +P_{[100]}$ (\textbf{F}) domain boundaries.
    \textbf{G}, Magnon transport hysteresis loops controlled via the gate electric field in source left drain right (sLdR) configuration. The domain between source and drain is polarized into $+P_{[100]}$ (red) and $-P_{[100]}$ (blue) states.
    \textbf{H}, Gate-controlled switching between the 'off' state (single domain) and 'on' state (domain boundary). 
    \textbf{I}, Reversible 'on' and 'off' switching under successive gate-field pulses of $\pm$ 200 kV/cm.
    }
	\label{fig:fig1}
\end{figure}

\paragraph{Gate-controlled magnon transistor.}
In the bulk, BFO crystallizes in the rhombohedral $R3c$ structure and exhibits robust ferroelectricity and G-type antiferromagnetic (AFM) order at room temperature \cite{lebeugle_roomtemperature_2007, catalan_physics_2009}. The magnetic structure consists of two antiparallel spin sublattices, $m_1$ and $m_2$ (Fig.\ref{fig:fig1}A). A Dzyaloshinskii--Moriya interaction (DMI), allowed by broken inversion symmetry, stabilizes a long-wavelength $\approx$ 62 nm cycloidal modulation of the underlying G-type AFM order, with a propagation direction coupled to the ferroelectric polarization \cite{sosnowska_spiral_1982,rahmedov_magnetic_2012, gross_realspace_2017, meisenheimer_switching_2024}. Recent studies have shown that rare-earth substitution promotes enhanced magnon transport at relatively smaller switching electric fields \cite{husain_nonvolatile_2024}. Here, we employ BFO films with 15\% samarium (Sm) substitution (i.e., Bi$_{0.85}$Sm$_{0.15}$FeO$_3$, henceforth sBFO). Unlike charge transport in metals, spin transport in insulators such as BFO occurs through propagating magnons.

To investigate gate-controlled magnon transport, we designed lateral devices based on an epitaxial sBFO film grown on a DyScO$_3$ (DSO) (110) substrate (see Supplementary Methods and Fig. \ref{fig:figS1}). Spin-orbit coupled platinum (Pt) electrodes serve as the magnon source and detector, while an independent gate electrode controls the antiferromagnetic domain configuration near the drain (Fig.\ref{fig:fig1}B). Electric fields can be independently applied between the source and drain ($E_{SD}$) and between the gate and drain ($E_{G}$). An electrical current through the Pt source generates a thermal gradient via Joule heating, driving magnons via spin Seebeck effect, as supported by the quadratic current dependence and second-harmonic response of the inverse-spin-Hall effect (ISHE) signal (Supplementary Fig. \ref{fig:figS2})\cite{uchida_observation_2008, uchida_spin_2010}. After propagating through the insulating sBFO channel, the non-equilibrium magnons transfer spin angular momentum into the Pt detector (the drain), generating a voltage through the ISHE \cite{guo_intrinsic_2008, sinova_spin_2015}. The ferroelectric polarization and nonlocal ISHE response exhibit correlated hysteresis upon sweeping $E_{SD}$ (Fig.\ref{fig:fig1}C), revealing a direct correlation between ferroelectric switching and magnon transport \cite{harris_symmetry_2025, husain_nonvolatile_2024}.

To identify the origin of the $V_{ISHE}$ hysteresis, we compare magnon transport through uniformly polarized channels and across boundaries separating oppositely polarized domains (Fig.\ref{fig:fig1}, D to F), with the source on the left and drain on the right (sLdR). The source-drain region was first polarized in the $+P_{[100]}$ (red) or $-P_{[100]}$ (blue) state by applying $E_{SD}$, followed by measurements of the ISHE response as a function of $E_{G}$ (Fig.\ref{fig:fig1}G). The two uniformly polarized configurations ($+P_{[100]} \rightarrow +P_{[100]}$ and $-P_{[100]} \rightarrow -P_{[100]}$) exhibit comparable output signals, indicating that the polarization of the source-drain region alone barely contributes to the observed hysteresis (see Supplementary Fig. \ref{fig:figS3} for domain imaging). In contrast, gated polarization reversal creates or removes an antiferromagnetic domain boundary near the drain, producing pronounced $V_{ISHE}$ hysteresis with opposite signs for the two domain-boundary configurations (see Supplementary Fig. \ref{fig:figS4} for details of the boundary origin). Specifically, $+P_{[100]} \rightarrow -P_{[100]}$ produces a positive ISHE signal, whereas $-P_{[100]} \rightarrow +P_{[100]}$ produces a negative signal, suggesting that reversing the domain-boundary orientation reverses the detected magnon spin polarization. We attribute this sign reversal to the odd-parity magnon spin polarization associated with cycloidal magnetic order, which will be discussed later. The domain-boundary-dependent ISHE response enables gate-controlled switching between low-signal 'off' states (uniform polarization) and high-signal 'on' states (opposite polarization, Fig.\ref{fig:fig1}H). Successive gate-field pulses of $\pm$ 200 kV/cm reversibly switch the device between 'on' and 'off' states (Fig.\ref{fig:fig1}I), demonstrating electrically programmable magnon transport.

\begin{figure}
	\centering
	\includegraphics[width=\textwidth]{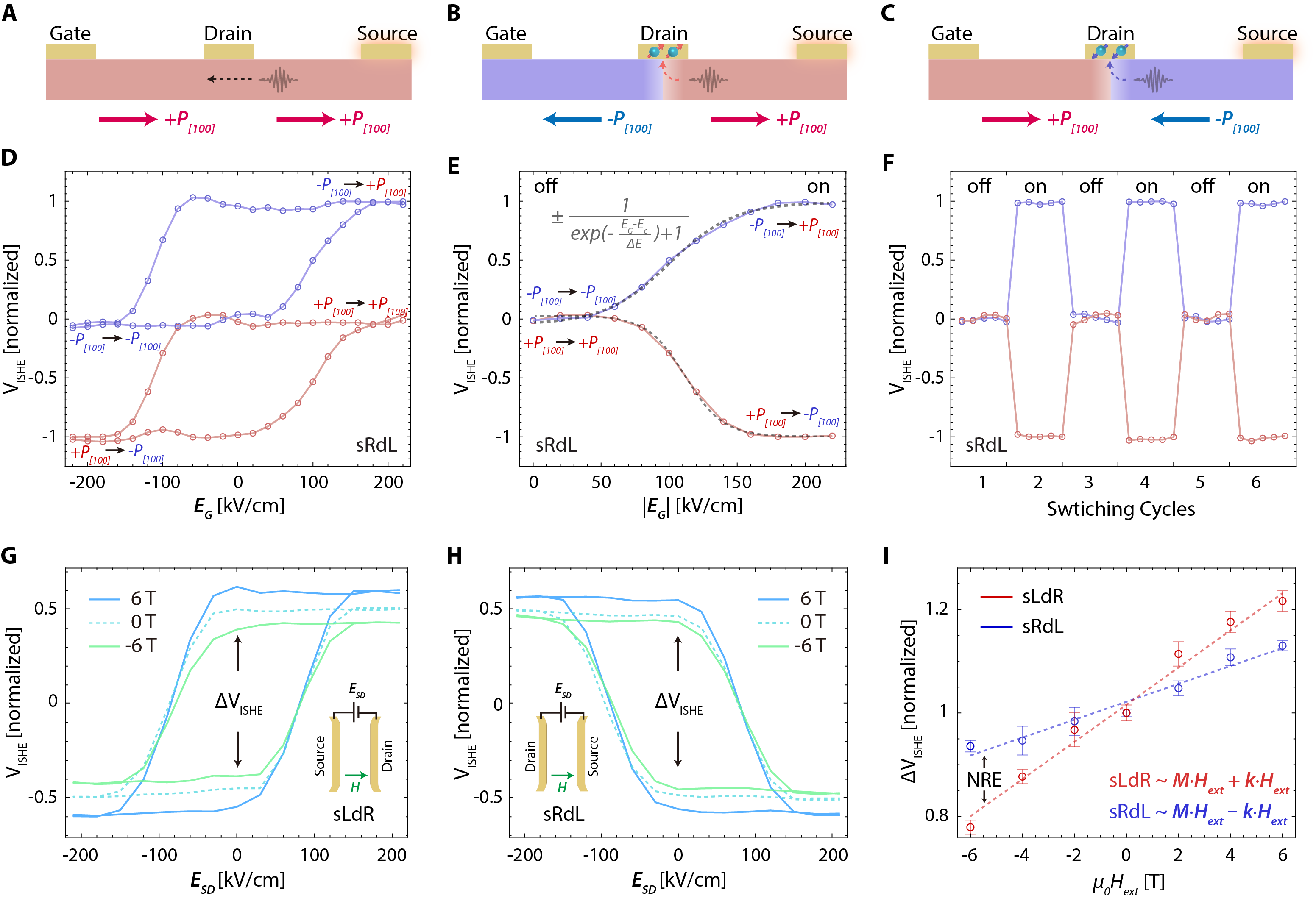}
	\caption{\textbf{Nonreciprocal magnon transport in BiFeO$_3$.}
    \textbf{A-C}, Schematics of right-to-left magnon propagation through $+P_{[100]} \rightarrow +P_{[100]}$ (\textbf{A}), $+P_{[100]} \rightarrow -P_{[100]}$ (\textbf{B}), and $-P_{[100]} \rightarrow +P_{[100]}$ (\textbf{C}) polarization configurations.
    \textbf{D}, Gate-dependent V$_{ISHE}$ hysteresis loops in the source-right, drain-left (sRdL) configuration, with the source-drain region polarized into $+P_{[100]}$ (red) or $-P_{[100]}$ (blue).
    \textbf{E}, Gate-controlled switching between the low-signal 'off' state (uniform polarization) and high-signal 'on' state (domain boundary). The data are fitted with a sigmoidal function (grey dashed line), where $E_c$ is the characteristic switching field and $\Delta E$ describes the switching width.
    \textbf{F}, Reversible 'on'/'off' switching under successive gate-field pulses of $\pm$200kV/cm.
    \textbf{G, H}, V$_{ISHE}$ hysteresis loops under magnetic fields of $-6$, $0$, and $6$ T for the sLdR (\textbf{G}) and sRdL (\textbf{H}) configurations. The V$_{ISHE}$ hysteresis amplitude is denoted by $\Delta$V$_{ISHE}$.
    \textbf{I}, Magnetic-field dependence of $\Delta$V$_{ISHE}$ for opposite magnon propagation directions. The difference in slopes indicates nonreciprocal magnon transport.
    }
	\label{fig:fig2}
\end{figure}

\paragraph{Nonreciprocal magnon transport.}
To understand the microscopic origin of the V$_{ISHE}$ hysteresis, we reversed the source, drain, and gate configuration, such that magnons propagated from right to left (sRdL), opposite to the sLdR configuration discussed above (Fig.\ref{fig:fig1}). The corresponding magnon propagation through uniformly polarized channels and across oppositely polarized domain boundaries is shown (Fig.\ref{fig:fig2}A-C). Notably, reversing the magnon propagation direction also reverses the sign of the gate-dependent V$_{ISHE}$ hysteresis (Fig. \ref{fig:fig2}D). In particular, the $+P_{[100]} \rightarrow -P_{[100]}$ domain boundary now produces a negative ISHE signal, whereas the $-P_{[100]} \rightarrow +P_{[100]}$ configuration yields a positive signal, opposite to the responses observed in the sLdR case (Fig.\ref{fig:fig1}G). Despite this reversal, the gate maintains control over magnon transport, switching the device between low-signal 'off' states with uniform ferroelectric polarization and high-signal 'on' states containing antiferromagnetic domain boundaries (Fig. 2E). The gate-dependent transition is well described by a sigmoidal function $V_{ISHE}=\pm V_0[1+\exp(-(E_G-E_c)/\Delta E)]^{-1}$, where the characteristic switching fields are $E_c=$102 and 114 kV/cm, and the switching widths are $\Delta E$ = 22.5 and 14 kV/cm for the blue and red curves, respectively. This sigmoidal response exhibits an exponential threshold-like turn-on characteristic reminiscent of conventional metal-oxide-semiconductor field-effect transistors (MOSFETs), although the underlying mechanism arises from ferroelectric domain reconfiguration rather than electrostatic carrier modulation. In conventional MOSFETs, thermally activated carrier transport imposes a fundamental limit on the subthreshold swing, constraining the gate-voltage change required to achieve a given On/Off current ratio and, consequently, the reduction of supply voltage and switching energy \cite{Sze2007,Zhirnov2003}. This limitation, commonly termed the “Boltzmann tyranny,” motivates the development of alternative switching mechanisms, including magnetoelectric devices. Repeated gate-field pulses (above saturation) reversibly configure the magnetic domains between these two states (Fig.\ref{fig:fig2}F). The reversal of the gate-dependent hysteresis upon reversal of the device orientation provides the first evidence of nonreciprocal magnon spin transport.


To further examine this directional asymmetry, we measured the ISHE response as a function of the source-drain electric field $E_{SD}$ for opposite magnon propagation directions. In the sLdR configuration, the ISHE exhibits a pronounced ferroelectric hysteresis with positive polarity (Fig.\ref{fig:fig2}G). Upon interchanging the source and drain electrodes (sRdL), the hysteresis polarity reverses (Fig.\ref{fig:fig2}H). Reversing the heater-detector configuration at fixed polarization reverses the sign of the nonlocal signal, demonstrating that the magnon spin signal is determined by the relative orientation of the propagation direction and domain boundary configuration, i.e., electrically-controlled non-reciprocal magnon transport.

Additional evidence emerges from the magnetic-field dependence of the ISHE response. We applied an external in-plane magnetic field ($\mu_0 H_{ext}$) along the magnon transport direction and measured the V$_{ISHE}$ hysteresis under $\mu_0 H_{ext}$ of $-6$, $0$, and $6$ T (Fig.\ref{fig:fig2}, G and H). The hysteresis amplitude, denoted by $\Delta$V$_{ISHE}$, increases approximately linearly with $\mu_0 H_{ext}$ for both propagation directions (Fig.\ref{fig:fig2}I). The $\mu_0 H_{ext}$ field dependence, however, exhibits distinctly different slopes for sLdR and sRdL. The common field-dependent contribution can be associated with the magnetic response of the canted antiferromagnetic order (denoted as $\mathbf{M}$), whereas the difference in slopes indicates an additional contribution proportional to $\mathbf{k}\cdot\mathbf{H_{ext}}$, that changes sign upon reversing the magnon propagation direction (denoted as $\mathbf{k}$), provided it is allowed by broken inversion and time-reversal symmetries of BFO \cite{tokura_nonreciprocal_2018}. Such a directional contribution is consistent with anti-symmetric spin splitting away from the $\Gamma$ point in a magnetic cycloid \cite{huang_manipulating_2024, brekke_minimal_2024, song_electrical_2025, yamada_metallic_2025}. These results reveal a pronounced nonreciprocal magnon transport across antiferromagnetic domain boundaries, providing the experimental basis for the microscopic mechanism discussed below.



\begin{figure}
	\centering
	\includegraphics[width=0.95\textwidth]{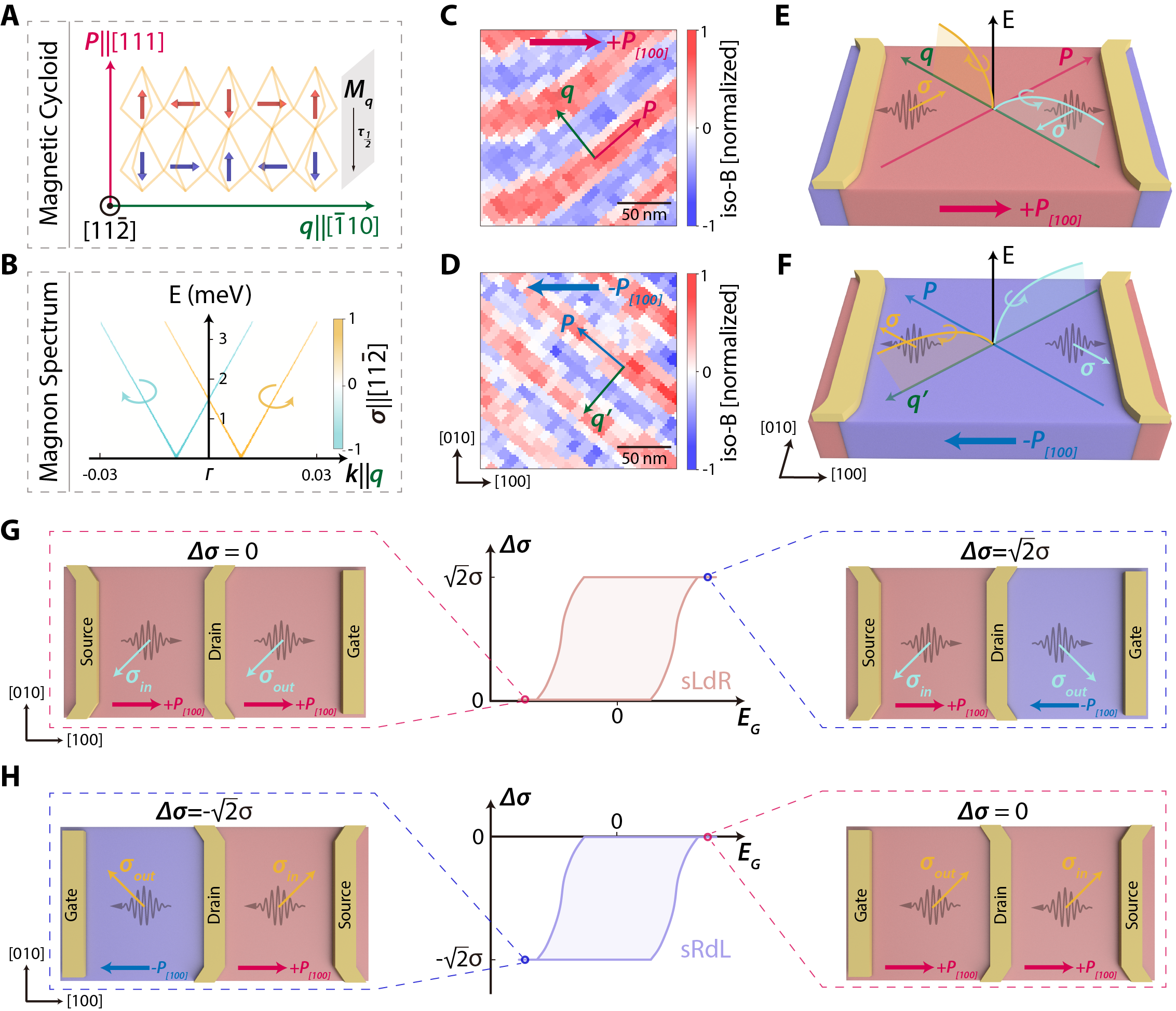}
	\caption{\textbf{Spin angular momentum transfer across antiferromagnetic domain boundaries.}
    \textbf{A}, The magnetic cycloid in BiFeO$_3$. The cycloid propagating along $q\parallel[\bar{1}10]$ preserves the combined mirror and half-translation symmetry ($M_q\tau_{1/2}$). 
    \textbf{B}, Magnon spectrum in BiFeO$_3$. The magnetic symmetry $M_q\tau_{1/2}$ protects odd-parity magnon spin splitting in a cycloid with magnon spin angular momentum ($\boldsymbol{\sigma}$) along $[11\bar{2}]$.
    \textbf{C,D}, Scanning nitrogen-vacancy microscopy of BiFeO$_3$ after poling in an electric field. The cycloid propagation vector $\mathbf{q}$ points along $[\bar{1}10]$ in $+P_{[100]}$ domain (\textbf{C}), and along $[\bar{1}\bar{1}0]$ in $-P_{[100]}$ domain (\textbf{D}).
    \textbf{E,F}, Schematic of the ferroelectric polarization ($\mathbf{P}$), magnon momentum ($\mathbf{k}$), and magnon spin angular momentum ($\boldsymbol{\sigma}$) in $+P_{[100]}$ (\textbf{E}), and $-P_{[100]}$ (\textbf{F}) domains.
    \textbf{G,H}, Gate-controlled magnon spin angular momentum transfer in the sLdR (\textbf{G}) and sRdL (\textbf{H}) configurations. Magnons crossing a domain boundary undergo a change in angular momentum, $\Delta\boldsymbol{\sigma}=\boldsymbol{\sigma}_{\mathrm{out}}-\boldsymbol{\sigma}_{\mathrm{in}}=\pm\sqrt{2}\sigma$, whose sign depends on the propagation direction and domain configuration.}
	\label{fig:fig3}
\end{figure}

\paragraph{Symmetry-ruled magnon transport.}
To elucidate the microscopic origin of nonreciprocal magnon transport in our field-effect magnon transistor, we perform a symmetry-based analysis of the coupling between ferroelectric polarization, cycloidal magnetic order, and magnon spin angular momentum in BFO. The magnetic ground state exhibits a long-wavelength cycloid propagating along $\mathbf{q}\parallel[\bar{1}10]$, with spin rotating in the $[\mathbf{P}, \mathbf{q}]$ plane (Fig.\ref{fig:fig3}A). Recent theoretical developments in $p$-wave magnetism suggest that cycloidal magnetic order can support odd-parity spin splitting, providing a microscopic basis for nonreciprocal magnon transport \cite{huang_manipulating_2024, brekke_minimal_2024, song_electrical_2025, yamada_metallic_2025, liang_ferroelectric_2026, shen_roomtemperature_2026}. The combined mirror and half-translation symmetry ($M_q\tau_{1/2}$) protects odd-parity magnon spin splitting, with the magnon spin angular momentum $\boldsymbol{\sigma}$ oriented along $[11\bar{2}]$. Consequently, magnons propagating in opposite directions along $\mathbf{q}$ carry opposite spin angular momenta, as illustrated by the magnon spectrum (Fig.\ref{fig:fig3}B). This momentum-dependent spin polarization provides the microscopic basis for nonreciprocal magnon transport. 



The strong coupling between ferroelectric and magnetic orders in BFO enables electrical control of the cycloidal spin texture. In particular, switching the in-plane ferroelectric polarization from $+P_{[100]}$ to $-P_{[100]}$ rotates the cycloidal propagation vector $\mathbf{q}$ by $90^\circ$ due to the locking of $\mathbf{P}$ and $\mathbf{q}$ via DMIs\cite{gross_realspace_2017, husain_nonvolatile_2024, meisenheimer_switching_2024}, as revealed by scanning nitrogen-vacancy microscopy (Fig.\ref{fig:fig3}C,D). This strong magnetoelectric coupling has been reported in both pure and rare-earth substituted BFO (Supplementary Fig. \ref{fig:figS5}). Since the magnon spin polarization is coupled to the cycloidal magnetic order, this reconfiguration also rotates the magnon spin angular momentum orientation (Fig.\ref{fig:fig3}E,F). In our device, the probed magnons propagate along [100], which forms a $45^\circ$ angle with the cycloidal wavevector $\mathbf{q}$, and therefore carry finite spin angular momenta. Consequently, electrically reconfiguring adjacent antiferromagnetic domains creates an interface between regions with distinct magnon spin polarizations.

When magnons traverse an antiferromagnetic domain boundary, their spin angular momentum can change as they interact with the local magnetic texture \cite{kim_propulsion_2014, lan_antiferromagnetic_2017, sayedaghaee_domainwall_2022}. This angular-momentum exchange modifies the spin polarization of the transmitted magnon current, thereby altering the spin current injected into the Pt drain and the resulting $V_{ISHE}$. We characterize this change by $\Delta\boldsymbol{\sigma}=\boldsymbol{\sigma}_{\mathrm{out}}-\boldsymbol{\sigma}_{\mathrm{in}}$, where $\boldsymbol{\sigma}_{\mathrm{in}}$ and $\boldsymbol{\sigma}_{\mathrm{out}}$ denote the incoming and outgoing magnon spin polarizations at the domain boundary, respectively. For oppositely polarized ferroelectric domains with orthogonal magnon spin polarizations of equal magnitude $\sigma$, the angular-momentum difference is $\Delta\boldsymbol{\sigma}=\pm\sqrt{2}\sigma$, where the sign depends on the magnon propagation direction and domain-boundary orientation. In contrast, a uniformly polarized channel preserves the magnon spin polarization, yielding $\Delta\boldsymbol{\sigma}=0$.

This picture provides a microscopic framework for gate-controlled magnon transport in this designed magnon transistor device (Fig.\ref{fig:fig3}G,H). For left-to-right propagation (sLdR), a uniformly polarized channel produces negligible magnon spin angular momentum change, whereas the formation of a gate-induced antiferromagnetic domain boundary produces a positive contribution to the detected spin signal (Fig.\ref{fig:fig3}G). Reversing the propagation direction (sRdL) reverses the sign of $\Delta\boldsymbol{\sigma}$, resulting in polarity inversion of the measured signal (Fig.\ref{fig:fig3}H). The opposite signs of the gate-dependent responses reveal the nonreciprocal magnon transport across antiferromagnetic domain boundaries (see Supplementary Fig. \ref{fig:figS6} for further analysis). Thus, electrically programmable domain boundaries enable directional control of magnon spin transport and the resulting ISHE response, providing the microscopic basis for the field-effect magnon transistor.


\paragraph{First-principles calculations and magnon transistor simulations.}

To explore the potential of field-effect magnon transistors for information processing, we performed circuit-level simulations using Verilog-A and Cadence Spectre (Supplementary Fig.\ref{fig:figS7} and \ref{fig:figS8}). The model incorporates two independently programmable ferroelectric regions, a magnon excitation source, and an electronic readout circuit. Binary inputs are encoded in the polarization states of the two regions, which remain unchanged after electrical programming. Based on the experimentally observed domain-dependent ISHE response, the model assumes that uniformly polarized configurations produce negligible voltage, whereas oppositely polarized configurations generate signals of opposite signs. A voltage receiver converts these analog signals into digital outputs, assigning logical 1 to matching polarization states and logical 0 to opposite states, thereby implementing an exclusive-NOR (XNOR) operation. Transient simulations reproduce the expected output sequence of 1, 0, 0, and 1 for the four input combinations (00, 01, 10, and 11), respectively (Supplementary Fig.\ref{fig:figS9}). The ISHE signals are converted into digital voltage levels using an externally powered receiver. These simulations demonstrate the circuit-level feasibility of integrating nonvolatile multiferroic state programming with magnon-based logic operations, providing a potential route toward electrically programmable magnonic information processing.


\paragraph{Discussion and outlook.}
Our demonstration of a field-effect magnon transistor provides a new route toward electrically programmable spin-based information processing. The device exhibits pronounced nonlinear switching, with a sigmoidal dependence on the gate field that resembles the turn-on characteristics of conventional field-effect transistors. Beyond its technological potential, our results provide fundamental insight into magnon transport in noncollinear antiferromagnets. By combining complementary magnon hysteresis measurements with symmetry analysis and first-principles calculations, we identify a mechanism for nonreciprocal magnon transport rooted in cycloidal magnetic order. Unlike mechanisms relying on weak canted moments in collinear antiferromagnets, which can produce magnon spin splitting at the Brillouin-zone center, cycloidal magnetic order enables momentum-dependent magnon spin polarization in a nearly-compensated magnet, with spin splitting emerging at finite wavevectors—a concept closely related to recent developments in altermagnetism and kinetomagnetism\cite{smejkal_chiral_2023, liu_chiral_2024, sun_observation_2025, liu_switchable_2026, cheong2024kinetomagnetism,1wyj-nrgk}. Antiferromagnetic domain boundaries then provide interfaces for propagating magnons to exchange spin angular momentum with the local magnetic texture, thereby amplifying/modulating magnon accumulation at the detector. It is noteworthy that the effects reported in this article can be significantly enhanced through the introduction of spin-orbit coupled metals with a much larger spin-Hall angle as compared to Pt (e.g., SrIrO$_3$ is $\approx$ 15-times larger than Pt \cite{patri_theory_2018, everhardt_tunable_2019, huang_manipulating_2024};  topological insulators can likely exhibit even higher spin-Hall angles and thus a correspondingly higher inverse-spin-Hall voltage in such a spin transistor\cite{nan_anisotropic_2019, mellnik_spin_2014, jiang_enhanced_2016, wang_surface_2016}. Finally, our model geometry is in the lateral direction (similar to a CMOS transistor);  evolving into a vertical geometry will significantly reduce the voltage required to program such a transistor, but will require further innovations to architect the domain boundary in such a geometry. 


\clearpage
\bibliography{BFO.bib}
\bibliographystyle{sciencemag}


\section*{Acknowledgments}
\paragraph*{Funding:}
Q.S. acknowledges that this research was sponsored by the Army Research Laboratory and was accomplished under Cooperative Agreement Number W911NF-24-2-0100. The views and conclusions contained in this document are those of the authors and should not be interpreted as representing the official policies, either expressed or implied, of the Army Research Laboratory or the U.S. Government. The U.S. Government is authorized to reproduce and distribute reprints for Government purposes notwithstanding any copyright notation herein. This work is partially supported by the Defense Advanced Research Projects Agency (DARPA) under Agreement No. HR0011269E210, No. HR0011269E209, and HR0011269E203 as part of the "Fast and Curious program". The U.S. Government is authorized to reproduce and distribute reprints for Government purposes notwithstanding any copyright notation herein. S.P., L.B., L.W.M., D.G.S., M.R., and R.R. also acknowledge support from the Army Research Office under the ETHOS MURI via cooperative agreement W911NF-21-2-0162. S.P. and L.B. additionally acknowledge support from the Vannevar Bush Faculty Fellowship (VBFF) from the Department of Defense and an Impact Grant 3.0 from ARA.
\paragraph*{Author contributions:}
Q.S. and R.R. conceived the project. Q.S., P.G., M.R. and S.P. synthesized and characterized the BFO film. A.O. fabricated the device. Q.S., S.H. and N.R. performed spin-transport measurements. P.M. and S.J. performed Nitrogen-Vacancy Microscopy measurements. Q.S. performed the symmetry analysis. S.P. and T.C. performed the density functional theory calculations and analysis. Z.Y. performed the transistor circuit simulation. B.H., Z.Z. and T.Z. performed magnon spectrum calculation. Q.S., S.H., L.M. and R.R. wrote the paper, with contributions from all co-authors.
\paragraph*{Competing interests:}
There are no competing interests to declare.
\paragraph*{Data, code and materials availability:}
The datasets generated and/or analyzed during the present study are available from the corresponding author on reasonable request.


\subsection*{Supplementary materials}
Materials and Methods\\
Supplementary Text\\
Figures S1 to S9\\
Tables S1 to S3


\newpage


\renewcommand{\thefigure}{S\arabic{figure}}
\renewcommand{\thetable}{S\arabic{table}}
\renewcommand{\theequation}{S\arabic{equation}}
\renewcommand{\thepage}{S\arabic{page}}
\setcounter{figure}{0}
\setcounter{table}{0}
\setcounter{equation}{0}
\setcounter{page}{1} 


\begin{center}
\section*{Supplementary Materials for\\ \scititle}

	Qian Song$^{1,2,3\ast}$,
    Sergei Prokhorenko$^{4}$,
    Thomas Cleveland$^{4}$,
    Ashish Omar$^{1}$,\\
    Zhi Yao$^{5}$,
    Sajid Husain$^{1,2}$,
    Peter Meisenheimer$^{1}$,
    Bowen Hao$^{6,7}$,\\
    Ziye Zhu$^{6,7}$,
    Pushpendra Gupta$^{1}$,
    Shashank Ojha$^{3,8}$,
    Nicholas Reiterer$^{1}$,\\
    Maya Ramesh$^{9}$,
    Sergio Puebla$^{1}$,
    Darrell Schlom$^{9}$,
    Tong Zhou$^{5,6}$,\\
    Laurent Bellaiche$^{4,10}$,
    Lane W. Martin$^{3,8,11}$,
	Ramamoorthy Ramesh$^{1,2\ast}$\\
\small$^\ast$Corresponding author. Email: qiansong@berkeley.edu, rramesh@berkeley.edu\\
\end{center}

\subsubsection*{This PDF file includes:}
Materials and Methods\\
Supplementary Text\\
Figures S1 to S9\\
Tables S1 to S3

\newpage


\subsection*{Materials and Methods}
\noindent
\paragraph{Thin film deposition.}
Sm-doped BiFeO$_3$ (15\% Sm) thin films were grown on DyScO$_3$ (110) substrates by pulsed laser deposition (PLD) using a KrF excimer laser (248 nm, COMPex-Pro, Coherent). The target-to-substrate distance was maintained at 50 mm, and the film thickness was 90 nm unless otherwise specified. Before deposition, the substrates were cleaned with isopropyl alcohol (IPA) and acetone for 5 min each and mounted on the substrate heater using silver paint to ensure good thermal contact. Deposition was performed at 710 $^\circ$C under a dynamic oxygen pressure of 140 mTorr, with a laser fluence of 1.8 J/cm$^2$ and a repetition rate of 15 Hz. After deposition, the films were cooled to room temperature at 30 $^\circ$C/min in oxygen atmosphere at 400 Torr pressure. The crystalline quality and surface morphology of the films were characterized by X-ray diffraction (X'Pert MRD, PANalytical) and atomic force microscopy (MFP-3D, Asylum Research), respectively.

\paragraph{Piezoelectric force microscopy.}
Piezoresponse force microscopy (PFM) measurements were performed using an MFP-3D atomic force microscope (Asylum Research) operated in dual AC resonance tracking (DART) mode. Lateral PFM was employed to characterize the in-plane piezoresponse and ferroelectric domain configurations. A Pt-coated silicon cantilever was used as the conductive probe for imaging and local electric-field application.

\paragraph{Device fabrication.}
Device fabrication began with ultrasonic cleaning in acetone and IPA. The sBFO film was transferred to a high-vacuum DC magnetron sputtering chamber for Pt deposition. A 15-nm-thick Pt layer was deposited at room temperature under an Ar pressure of 7 mTorr and a sputtering power of 15 W. A 500-nm-thick layer of positive photoresist (MIR 701) was spin-coated at 7000 rpm for 60 s, followed by baking at 100 $^\circ$C for 60 s. Photolithography was performed using a mask aligner (MA6, Karl Suss) with i-line exposure at an intensity of 10 mW/cm$^2$ for 5 s. The exposed photoresist was developed in MEGAPOSIT MF-26A for 20 s. The patterned Pt layer was subsequently defined by ion milling (Nanoquest, Intlvac, equipped with a Hiden Analytical secondary ion mass spectrometry system) down to the sBFO surface, forming rectangular electrodes with dimensions of 120 $\mu$m $\times$ 1.3 $\mu$m. Device fabrication was carried out at the Marvell Nanofabrication Laboratory, University of California, Berkeley.

\paragraph{Nonlocal magnon transport measurements.}
Nonlocal magnon transport measurements were performed using six-terminal devices (Supplementary Fig. \ref{fig:figS4}A), with independently addressable left and right gate electrodes for controlling antiferromagnetic domain boundaries. An alternating current (at 13 Hz) was applied to the Pt source electrode using a Keithley 6200 current source to generate thermally driven magnons via the spin Seebeck effect. The nonlocal inverse spin Hall effect (ISHE) voltage was measured using an SR830 lock-in amplifier referenced to the second harmonic of the current source frequency. A typical value of V$_{ISHE}$ hysteresis amplitude is 0.6 $\mu$V (at 3 mA current supply), and it is normalized in this study. Measurements and electrical switching between different electrode configurations were automated using custom Python software and a Keithley 7001 switch system.

\paragraph{Scanning nitrogen-vacancy microscopy.}
Scanning nitrogen-vacancy (NV) microscopy of BFO samples were measured at room temperature using a commercial NV microscope (Qnami ProteusQ), which combines a confocal optical microscope with a tuning-fork based atomic force microscope. Diamond tips with a parabolic taper containing single NV centers were used to increase photon collection efficiency (Quantilever MX+).




\subsection*{Supplementary Text}
\subsubsection*{Experimental Section}
\paragraph{Thin film characterization.} The structural and ferroelectric properties of the 15\% Sm-doped BiFeO$_3$ thin films grown on DyScO$_3$ (110) substrates were characterized using atomic force microscopy (AFM), piezoresponse force microscopy (PFM), and high-resolution X-ray diffraction (XRD). The AFM topographic image (fig. S1A) reveals a continuous film surface with atomically step features. Lateral PFM images reveals large ferroelectric domains with opposite in-plane polarization contrast, together with narrower stripe-like domain patterns (fig. S1B). These features are consistent with the coexistence of distinct ferroelastic domain configurations in rhombohedral-like BFO. The XRD $\theta–2\theta$ scan (fig. S1C) displays well-defined diffraction peaks of BFO, consistent with crystalline thin-film growth. Reciprocal-space mapping (RSM) around an asymmetric Bragg reflection (fig. S1D) further supports the epitaxial ordering of the film. Together, these measurements establish the crystalline quality and ferroelectric domain structure of the BFO thin film, providing the material foundation for electrically controlled domain configurations and magnon transport investigated in this work.

\paragraph{Magnon transport.} To investigate the origin of the magnon spin current, we performed current-dependent and harmonic-resolved measurements of the nonlocal inverse spin Hall effect (ISHE) signal. Figure S2a shows an optical image of the device and the corresponding measurement configuration. An alternating current applied to the Pt source generates Joule heating, producing a thermal gradient that can excite magnons through the spin Seebeck effect. The resulting magnon spin current propagates through the BFO channel and is electrically detected at the Pt drain through the inverse spin Hall effect.

The hysteresis amplitude $\Delta V_\mathrm{ISHE}$, exhibits a quadratic dependence on the applied source current, $(\Delta\mathrm{ISHE}\propto I^2)$ (fig. S2B), consistent with a Joule-heating-driven mechanism. To further distinguish thermal and electrical contributions, we measured the second-harmonic ISHE responses while sweeping the source–drain electric field $E_{SD}$. the in-phase second-harmonic signal remains close to zero (fig. S2C). In contrast, the out-of-phase second-harmonic response exhibits pronounced ferroelectric hysteresis, with a reversible modulation of the ISHE signal of approximately 0.6 $\mu$V (fig. S2D).

The quadratic current dependence and second-harmonic out-of-phase switching response are consistent with thermally generated magnon spin transport mediated by the spin Seebeck effect. These measurements further demonstrate that the ferroelectric configuration modulates the thermally driven magnon signal detected at the Pt drain.

\paragraph{Electric-field control of ferroelectric domain configurations.} To investigate the electric-field control of ferroelectric domain configurations, we performed lateral PFM measurements after local $\pm$ 200 kV/cm electrical poling of the BFO thin film. Figure S4 shows four distinct polarization configurations obtained by independently switching two selected regions. The initial $+P_{[100]}\rightarrow+P_{[100]}$ configuration exhibits similar LPFM contrast in both regions (fig. S3A). Reversing the polarization of either region produces the $-P_{[100]}\rightarrow+P_{[100]}$ and $+P_{[100]}\rightarrow-P_{[100]}$ configurations, respectively (fig. S3B,C), characterized by contrasting piezoresponse signals. Finally, switching between both regions produces the $-P_{[100]}\rightarrow-P_{[100]}$ configuration (fig. S3D).

These measurements demonstrate reversible and spatially selective control of the ferroelectric polarization in BFO. The ability to establish regions with either matching or opposing polarization provides an experimental basis for engineering antiferromagnetic domain boundaries and investigating their influence on magnon spin transport.

\paragraph{Gate-controlled magnon transport.} To investigate the spatial origin of the gate-controlled magnon response, we fabricated a device with two independently addressable gate electrodes positioned on opposite sides of the magnon transport channel (fig. S4A). This configuration allows the ferroelectric polarization to be locally controlled near either the source or the drain, while the magnon propagation direction can be reversed by interchanging the source and drain electrodes.

When the right gate is activated, pronounced hysteretic switching of the ISHE signal is observed in the sLdR configuration, where the drain is also located on the right side (fig. S4B). In contrast, the hysteresis is strongly suppressed in the reversed sRdL configuration, where the right gate is adjacent to the source. Conversely, activating the left gate produces pronounced hysteresis in the sRdL configuration, while the sLdR configuration exhibits negligible switching (fig. S4C).

These measurements demonstrate that the gate-controlled magnon response depends critically on the relative positions of the gate and drain electrodes. Pronounced switching occurs only when the gate is positioned near the drain, supporting a mechanism in which the local antiferromagnetic domain boundary modifies the spin angular momentum reaching the Pt detector. The spatial selectivity further indicates that the observed modulation cannot be explained solely by a uniform change in magnon transport throughout the channel.

\paragraph{Comparison of pristine and rare-earth substituted BFO.} To investigate the influence of magnetic domain configurations on magnon spin transport, we compare the cycloidal domain structures and nonlocal ISHE responses of pristine BFO and 15\%Lanthanum-substituted BFO (La-BFO). Scanning nitrogen-vacancy magnetometry of pristine BFO reveals multiple cycloidal domains following electric-field poling, as reported previously \cite{meisenheimer_switching_2024}. In the $+P_{[100]}$ state (fig. S5A), the in-plane polarization $\mathbf{P}$ lies along [110] or $[1\bar{1}0]$, with the cycloidal wavevector $\mathbf{q}$ oriented perpendicular to $\mathbf{P}$. Reversing the polarization to the $-P_{[100]}$ state rotates both the polarization and cycloidal wavevector by $90^\circ$ (fig. S5B). The corresponding magnon transport measurements exhibit a relatively weak hysteretic modulation, $\Delta V_{{ISHE}}\approx150$ nV (fig. S5C), as reported previously \cite{harris_symmetry_2025}.

In contrast, La-BFO exhibits substantially larger, more uniformly oriented cycloidal domains following electric-field poling (fig. S5D,E) \cite{husain_nonvolatile_2024}. The corresponding magnon transport hysteresis reaches $\Delta V_{\mathrm{ISHE}}\approx500$ nV (fig. S5F), comparable to the response observed in the sBFO devices investigated in this work. These observations are consistent with the picture in which the cycloidal domain configuration accounts for the net spin polarization of the detected spin current. In pristine BFO, contributions from multiple cycloidal domains partially cancel, reducing the detected ISHE signal. By contrast, larger and more uniformly oriented cycloidal domains can reduce such cancellation and enhance the net spin signal. This comparison further supports the role of cycloidal magnetic order and its associated magnon spin polarization in the electrically controlled ISHE response.

\paragraph{Symmetry-based model of spin angular momentum transfer at the antiferromagnetic domain boundary.} To elucidate the microscopic origin of the gate-controlled magnon response, we consider the spin angular momentum exchange between propagating magnons and the local magnetic texture at antiferromagnetic domain boundaries (fig. S6). In BFO cycloid phase, the magnon spin polarization is coupled to the cycloidal magnetic order, which can be reconfigured through ferroelectric switching. When magnons propagate across a boundary separating distinct antiferromagnetic domains, their spin polarization can change, resulting in angular momentum transfer to the magnetic texture. We characterize this change by $\Delta\boldsymbol{\sigma}=\boldsymbol{\sigma}_{\mathrm{out}}-\boldsymbol{\sigma}_{\mathrm{in}}$, where $\boldsymbol{\sigma}_{\mathrm{in}}$ and $\boldsymbol{\sigma}_{\mathrm{out}}$ denote the incoming and outgoing magnon spin polarizations across the domain boundary, respectively.

Figures S6A and S6B illustrate the proposed mechanism for gate-controlled switching in the two opposite source–drain configurations. In a uniformly polarized channel, the incoming and outgoing magnon spin polarizations remain unchanged, yielding $\Delta\boldsymbol{\sigma}=0$. In contrast, ferroelectric switching beneath the drain creates an antiferromagnetic domain boundary, across which the magnon spin polarization rotates. For a 90$^\circ$ rotation between spin-polarization vectors of equal magnitude $\sigma$, the angular momentum change has a magnitude $\Delta\boldsymbol{\sigma}=\sqrt{2}\sigma$. Reversing the magnon propagation direction reverses the signed angular momentum transfer in this model, accounting for the opposite gate-dependent switching responses in the sLdR and sRdL configurations.

We further consider the effect of the source–drain electric field $E_{SD}$ (fig. S6C,D). In this configuration, the relevant quantity is the projection of the magnon spin angular momentum change along the [100] direction. For the spin orientations illustrated, the projected change takes values of $\Delta\sigma_{[100]}=\pm\sqrt{2}\sigma/2$, depending on the ferroelectric configuration and magnon propagation direction. These configurations provide a phenomenological picture for the reversal of the hysteretic ISHE response upon interchanging the source and drain electrodes.

Together, these schematics illustrate how the spatial arrangement of ferroelectric domains and the direction of magnon propagation can determine the sign and magnitude of the magnon spin angular momentum exchanged at an antiferromagnetic domain boundary, thereby modifying the spin polarization of the magnon current reaching the Pt detector.

\subsubsection*{Circuit model operation and logic mapping.}
The simulation tests a domain-state comparison followed by electrical readout: two locally programmed states are retained, a separate read excitation produces an analogue contrast signal, and a powered receiver reports whether the states match. The input bits $A$ and $B$ are encoded in the retained states; they are not the read-source voltage or the receiver supply.

\textbf{Write the inputs.} Independent pulses WA and WB set the two state variables $s_A,s_B\in\{+1,-1\}$. In this encoding, bit 0 corresponds to $+1$ and bit 1 to $-1$. Removing a write pulse leaves the selected state unchanged in the model. These signs label opposite programmed states, rather than high and low detector voltages.

\textbf{Read the contrast.} In Fig. \ref{fig:figS7}, SRC supplies the same read excitation for all four input combinations. The assumed detector response is proportional to the difference between the two states: equal states give approximately zero raw voltage, whereas unequal states give opposite signed voltages. At the nominal settled read drive, the unequal-state levels are +0.6 and -0.6 $\mu$V. Both are \emph{large-magnitude} signals; a negative raw voltage does not by itself mean logic 0.

\textbf{Generate the logic output.} The receiver observes VRAW. During an enabled read, it drives Q high when $|V_{\mathrm{raw}}|<0.3$ $\mu$V and low outside that window. Thus the raw-signal contrast and the digital output have the following mapping:
\begin{center}\small
\begin{tabular}{@{}lll@{}}\toprule
Programmed states & Raw-signal magnitude & Receiver output Q\\\midrule
Same ($00$ or $11$) & Low (approximately zero) & High: match\\
Different ($01$ or $10$) & High (approximately $0.6\mu$V) & Low: mismatch\\\bottomrule
\end{tabular}
\end{center}
This fixed mapping produces an XNOR output. The 1.8 V receiver supply provides output restoration. Q is valid only after the read excitation has settled and read enable is high; the raw signal is also near zero without excitation.

The three Verilog-A blocks implement the retained-state detector, powered receiver and stimulus generator. The following sections describe the wiring and sequence first, then the equations implementing each operation, the exported results and the model limits.

\paragraph{Cadence setup and stimulus.}
The library is \texttt{BFO\_XNOR\_DEMO}. The composite schematic \texttt{bfo\_xnor\_unit} connects domain-pair \texttt{vraw} to receiver \texttt{vin}; it exposes WA, WB and SRC for the detector, RD\_EN for receiver \texttt{en}, VDD for receiver \texttt{vdd}, and Q from receiver \texttt{q}. PA, PB and VRAW remain available as monitor nets. All ground terminals share one reference. Fig. \ref{fig:fig4} and Fig. \ref{fig:figS8} shows the testbench and the interfaces, respectively. Wiring follows terminal names: the leaf domain-pair symbol places \texttt{wb} above \texttt{wa}, whereas the wrapper places WA above WB.

\textbf{Simulation configuration and stimulus.}
The three leaf cells use their \texttt{veriloga} views. 
The composite unit and testbench use connected \texttt{schematic} views. 
The supplied standalone deck \texttt{xnor\_transient.scs} specifies a 400 $\mu$s transient at 27 $^\circ$C, with \texttt{maxstep=1u}, \texttt{errpreset=conservative}, \texttt{reltol=1e-5}, \texttt{vabstol=1e-9} and \texttt{iabstol=1e-15}. These are reproduction-deck settings; the exported CSVs do not record all options and instance overrides of the original server run.

The stimulus cycles through 00, 01, 10 and 11 in four 100 $\mu$s slots. Within each slot, the sequence is:
\begin{center}\small
\begin{tabular}{@{}ll@{}}\toprule
Time relative to slot start ($\mu$s)&Command\\\midrule
10--20&WA/WB = $+40$ V for bit 0, $-40$ V for bit 1\\
40--90&SRC = 3 V\\
60--85&RD\_EN = 1.8 V\\
75&Sample PA, PB, VRAW and Q\\\bottomrule
\end{tabular}
\end{center}
All other stimulus levels are zero, with 100 ns edge smoothing. Absolute sample times are 75, 175, 275 and 375 $\mu$s. Writes have ended before each read. Enable rises ten read time constants after source turn-on, so that the receiver evaluates the settled contrast rather than the near-zero signal before read excitation. The first 00 slot starts from initialized $+1,+1$ states; subsequent slots exercise state reversals.

\paragraph{Model equations.} The following equations implement the write, read and receiver operations described above. Full parameter defaults are listed in the final section.

\textbf{Programming and retained states.}
The write operation updates each region independently. For $j\in\{A,B\}$, $W_j$ is the local write voltage and $V_{c,j}$ is the positive switching-threshold magnitude:
\begin{equation}
s_j\leftarrow\begin{cases}
+1,&W_j\text{ crosses }+V_{c,j}\text{ upwards},\\
-1,&W_j\text{ crosses }-V_{c,j}\text{ downwards},\\
s_j,&\text{otherwise}.
\end{cases}\label{eq:state}
\end{equation}
Here ''upwards'' and ''downwards'' refer to the direction of voltage crossing, not to a spatial polarization direction. The otherwise branch retains the previous state. This rule selects the states; the detector equation below determines their readout contrast.

Both thresholds have magnitude $V_{c,j}=20$ V. Initialization sets $s_j=1-2\,\texttt{bit\_j\_init}$, with an override when the initial write voltage exceeds a threshold. The default initial bits are 00. Positive and negative 40 V write pulses select bit 0 ($s_j=+1$) and bit 1 ($s_j=-1$), respectively. The last state is retained between switching events. The Verilog-A \texttt{transition} operator produces the normalized monitors $p_A$ and $p_B$ with 100 ns smoothing. PA and PB encode these dimensionless states as diagnostic voltages of approximately $\pm1$ V; they are not physical polarization voltages.

Each write terminal has a 1 T$\Omega$ leakage resistance and a 1 fF capacitance. These represent electrical loading only; polarization-switching loss and electrode capacitance require separate characterization. The write thresholds are voltage parameters. Converting 20 V into 100 kV cm$^{-1}$ would require an effective 2 $\mu$m field path, which is not established by this model.

\textbf{Read excitation and detector response.}
The source resistance is $R_{\mathrm{src}}=1$ k$\Omega$. A normalized heating drive and read envelope are defined by
\begin{equation}
I_{\mathrm{src}}=\frac{V_{\mathrm{SRC}}}{R_{\mathrm{src}}},\qquad
D=\left(\frac{I_{\mathrm{src}}}{I_{\mathrm{ref}}}\right)^2,\qquad
\tau_{\mathrm{read}}\frac{du}{dt}+u=D,
\label{eq:read}
\end{equation}
where $I_{\mathrm{ref}}=3$ mA and $\tau_{\mathrm{read}}=2\mu$s. At $V_{\mathrm{SRC}}=3$ V, the settled drive is unity. The square law represents normalized Joule-heating excitation. It does not calculate temperature or magnon populations. VRAW is a signed detector envelope corresponding to a spin-Seebeck excitation/ISHE readout interpretation; the model does not simulate the lock-in carrier or demodulation process.

The open-circuit detector voltage is
\begin{equation}
V_{\mathrm{open}}=V_{\mathrm{offset}}+\eta u\left[
V_{\mathrm{contrast}}\frac{p_A-p_B}{2}
+V_{\mathrm{common}}\frac{p_A+p_B}{2}\right].\label{eq:raw}
\end{equation}
The fixed measurement-polarity factor is $\eta=+1$. Nominally, $V_{\mathrm{contrast}}=0.6\mu V$ and $V_{\mathrm{common}}=V_{\mathrm{offset}}=0$. The difference term implements the equal-state cancellation introduced above; the common-state and offset terms allow departures from that idealization. These coefficients require calibration for the physical device and measurement geometry.

The detector has source resistance $R_{\mathrm{det}}=1$ k$\Omega$. Its Norton-equivalent contribution and loaded response are
\begin{align}
I(\mathrm{VRAW},\mathrm{gnd})&=\frac{V_{\mathrm{raw}}-V_{\mathrm{open}}}{R_{\mathrm{det}}},\\
C_{\mathrm{in}}\frac{dV_{\mathrm{raw}}}{dt}
+\frac{V_{\mathrm{raw}}}{R_{\mathrm{in}}}
+\frac{V_{\mathrm{raw}}-V_{\mathrm{open}}}{R_{\mathrm{det}}}&=0.
\end{align}
For the receiver input $R_{\mathrm{in}}=1$ T$\Omega$ and $C_{\mathrm{in}}=1$ pF, the DC loading factor is approximately $1-10^{-9}$ and the input RC time is approximately 1 ns. The visible microsecond rise and fall of VRAW therefore come from the assumed read-envelope time constant. A unit step gives $u=1-\exp[-(t-t_0)/\tau_{\mathrm{read}}]$, reaching 95\% in approximately 6 $\mu$s; removal of the drive gives exponential decay.

\textbf{Window receiver and output stage.}
The receiver uses VRAW, enable, supply and ground only. PA and PB are not receiver inputs. Its decision is
\begin{equation}
d=\begin{cases}
1,&V_{\mathrm{EN}}>0.9\,\mathrm{V}\text{ and }|V_{\mathrm{raw}}-V_{\mathrm{center}}|<\Delta,\\
0,&\text{otherwise},
\end{cases}
\qquad V_{\mathrm{center}}=0,\quad\Delta=0.3\mu \mathrm{V}.\label{eq:decision}
\end{equation}
A 100 ns transition produces $h(t)$ from $d$. Complementary output conductances drive the load:
\begin{align}
G_H&=R_{\mathrm{off}}^{-1}+(R_{\mathrm{on}}^{-1}-R_{\mathrm{off}}^{-1})h,\\
G_L&=R_{\mathrm{off}}^{-1}+(R_{\mathrm{on}}^{-1}-R_{\mathrm{off}}^{-1})(1-h),\\
C_L\frac{dV_Q}{dt}+G_H(V_Q-V_{\mathrm{DD}})+G_LV_Q&=0.
\end{align}
Here $R_{\mathrm{on}}=100\,\Omega$, $R_{\mathrm{off}}=1$ T$\Omega$, $C_L=10$ pF and $V_{\mathrm{DD}}=1.8$ V. VDD supplies output restoration. Q is forced low while read enable is low and is interpreted as a logic result only during an enabled read interval.

\paragraph{Exported transient and sampled results.}
The waveform export contains 2075 rows spanning 0--400 $\mu$s, with separate time/voltage columns for each of eight signals (Table \ref{tab:tabS3} and Fig. \ref{fig:fig4}). Each signal is linearly interpolated at the four sample times. Multiplication of VRAW by $10^6$ converts volts to microvolts; for example, \texttt{1e6*value(VT("/VRAW") 75e-6)} is the first ADE sample. All 28 corresponding scalar measurements agree with the separate ADE export within its displayed precision. No new Spectre run was used to generate these figures.

The equal-state samples fall inside the fixed window; both signs of the unequal-state response fall outside it. The resulting Q sequence is 1, 0, 0, 1 with unchanged read-source polarity and receiver settings. The exported high and low outputs are approximately 1.8 V and $1.8\times10^{-10}$ V, consistent with the finite on/off output conductances. The nominal distance from every sample to the relevant window boundary is approximately 0.3 $\mu $V.

\textbf{Parameters and model interfaces.}
The tables \ref{tab:tabS1}, \ref{tab:tabS2} list the supplied source defaults. The raw Verilog-A files, Spectre reproduction decks, CSV exports and plotting code accompany this supplement. Original run netlists and solver logs are needed to verify every setting of the originating server run.

\clearpage
\begin{figure}
	\centering
	\includegraphics[width=0.6\textwidth]{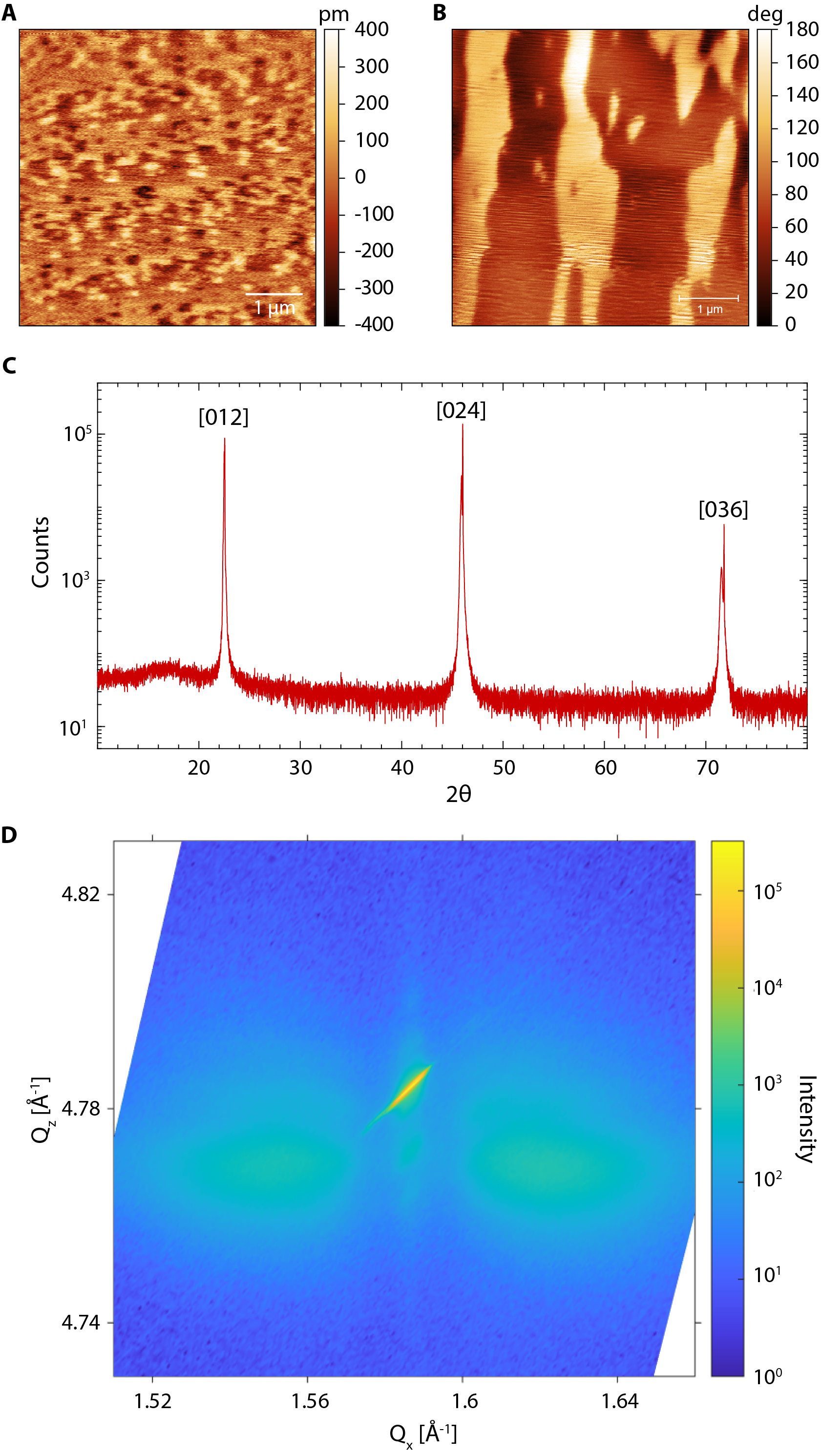} 
	\caption{\textbf{Structural and ferroelectric characterization of the BiFeO$_3$ thin film.}
    \textbf{A}, Atomic force microscopy topography of the grown film.
    \textbf{B}, Piezoresponse force microscopy image showing ferroelectric domains.
    \textbf{C}, X-ray diffraction pattern of the film.
    \textbf{D}, X-ray reciprocal space map of the film.
	}
	\label{fig:figS1}
\end{figure}

\clearpage
\begin{figure}
	\centering
	\includegraphics[width=0.7\textwidth]{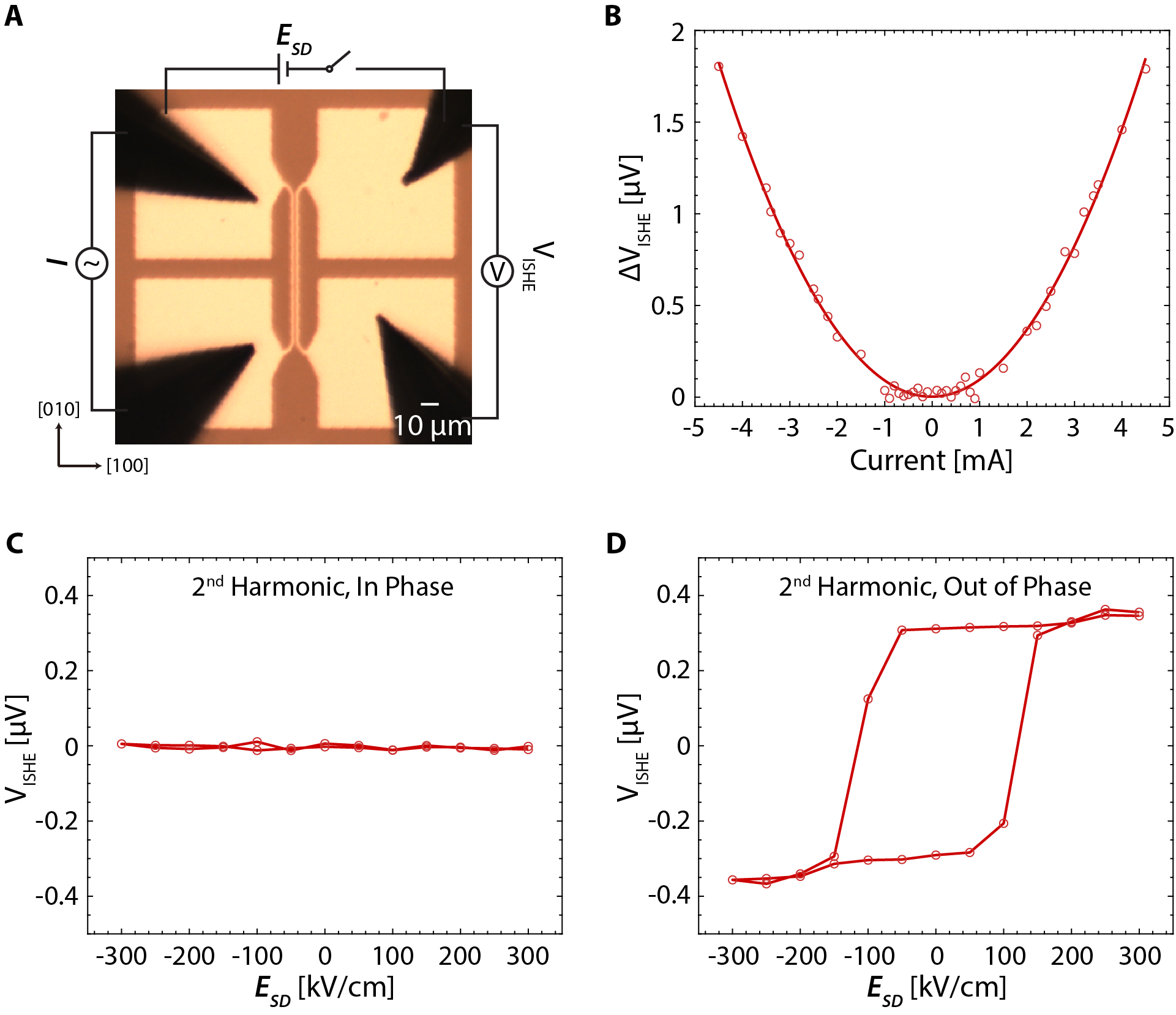} 
	\caption{\textbf{Spin Seebeck origin of magnon transport.}
    \textbf{A}, Optical image of the magnon transport device.
    \textbf{B}, Quadratic current dependence of the ISHE hysteresis amplitude ($\Delta V_{ISHE}$), consistent with thermally driven magnon transport via the spin Seebeck effect.
    \textbf{C,D}, Second-harmonic ISHE voltage as a function of the applied AC current. The pronounced hysteresis in the out-of-phase component supports the thermal origin of magnon transport.
}
	\label{fig:figS2}
\end{figure}

\clearpage
\begin{figure}
	\centering
	\includegraphics[width=0.7\textwidth]{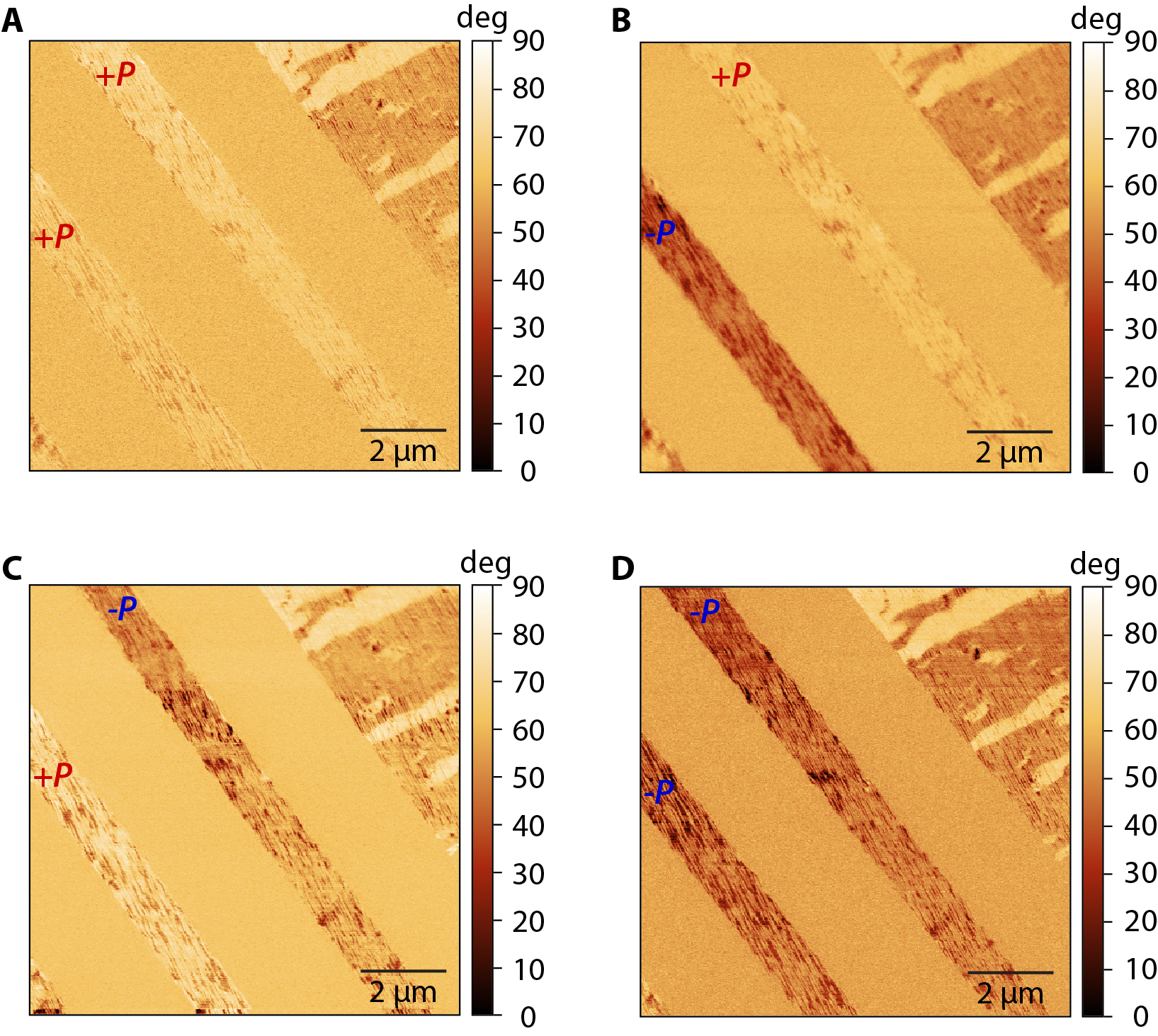} 
	\caption{\textbf{Ferroelectric domain switching in the field-effect magnon transistor.}
    Piezoresponse force microscopy (PFM) images showing four electrically programmed polarization configurations: \textbf{A}, $+P_{[100]} \rightarrow +P_{[100]}$; \textbf{B}, $-P_{[100]} \rightarrow +P_{[100]}$; \textbf{C}, $+P_{[100]} \rightarrow -P_{[100]}$; and \textbf{D}, $-P_{[100]} \rightarrow -P_{[100]}$. The device was rotated by 45$^\circ$ to visualize both ferroelectric domains and stripe-like ferroelastic domains.
    }
	\label{fig:figS3}
\end{figure}

\clearpage
\begin{figure}
	\centering
	\includegraphics[width=\textwidth]{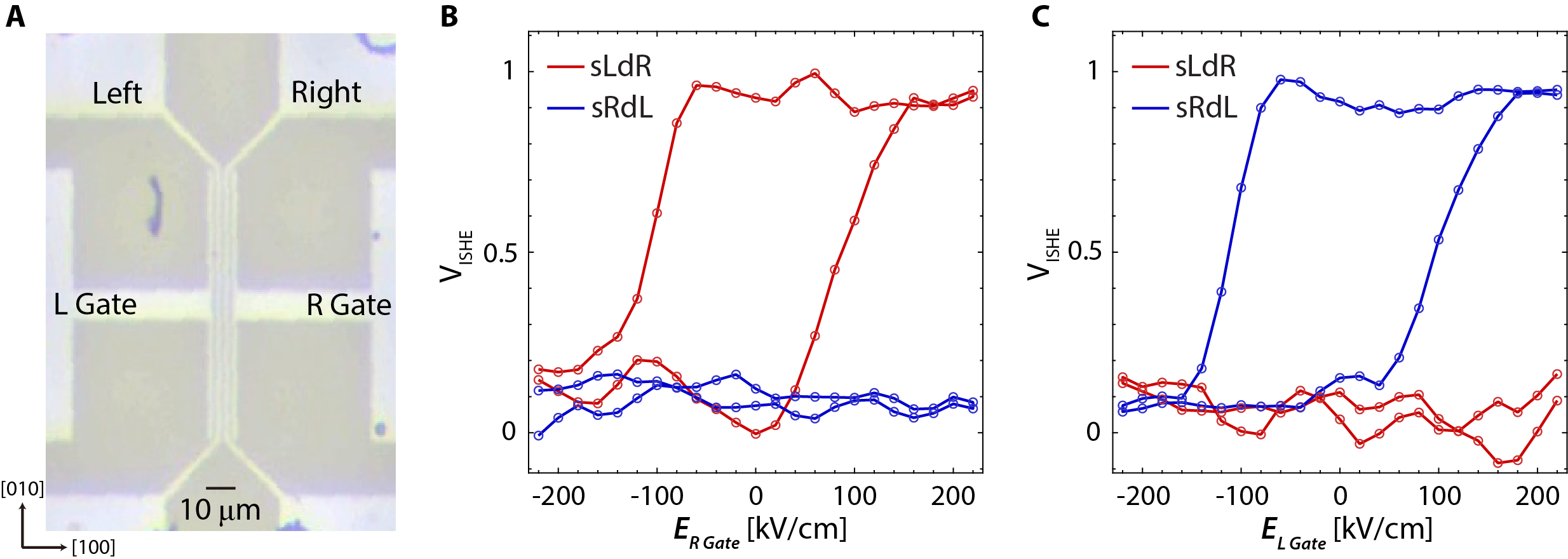} 
	\caption{\textbf{Domain-boundary origin of the V$_{ISHE}$ hysteresis.}
    \textbf{A}, Optical image of the field-effect magnon transistor.
    \textbf{B,C}, V$_{ISHE}$ hysteresis loops for the sLdR (red) and sRdL (blue) configurations as functions of the right (\textbf{B}) and left (\textbf{C}) gate electric field. Hysteresis emerges only when the gate and drain are on the same side, indicating that an antiferromagnetic domain boundary under the drain is responsible for the amplified magnon transport.
    }
	\label{fig:figS4}
\end{figure}

\clearpage
\begin{figure}
	\centering
	\includegraphics[width=\textwidth]{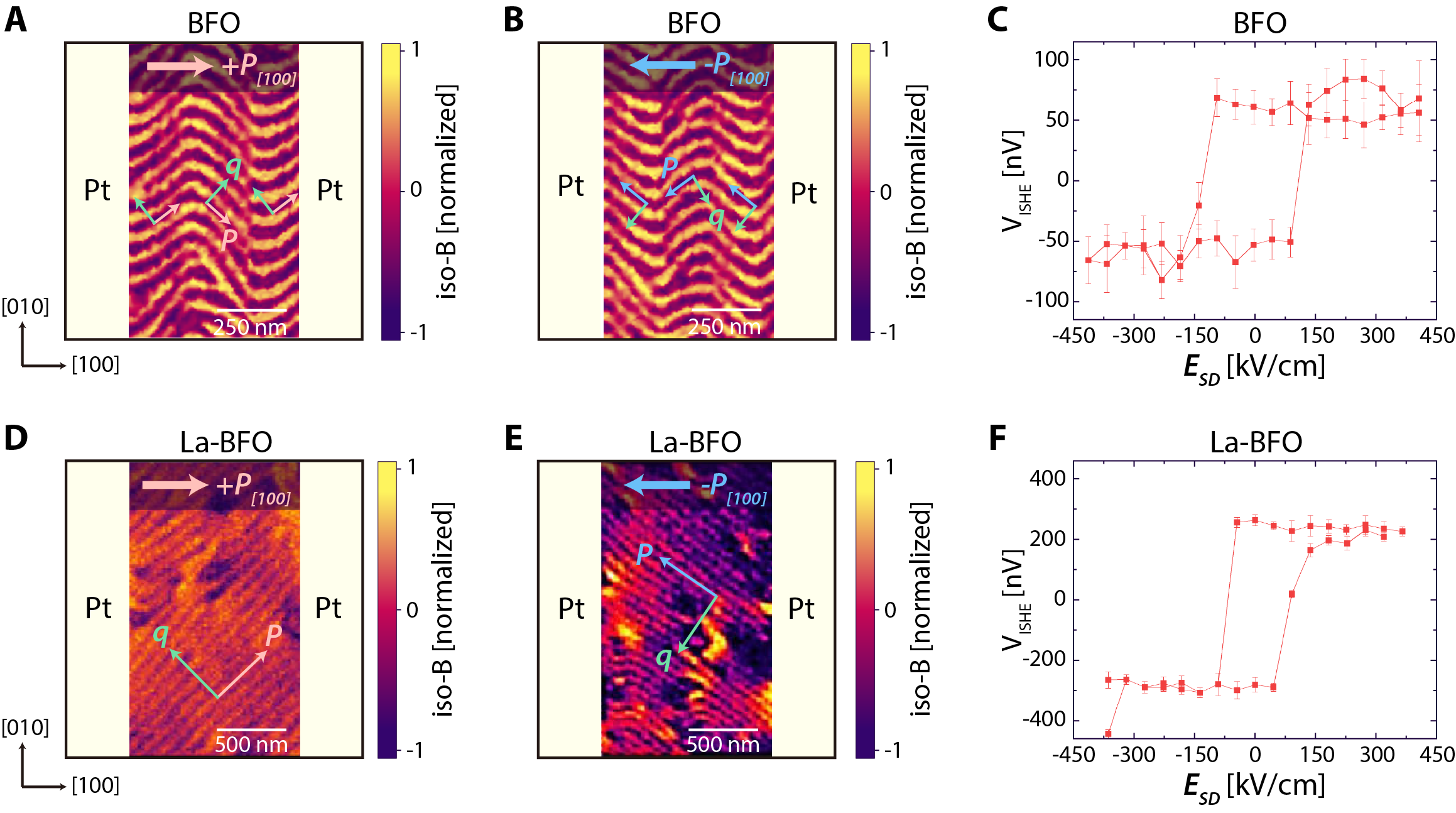} 
	\caption{\textbf{Large-scale nitrogen-vacancy microscopy and magnon transport of pristine and rare-earth substituted BiFeO$_3$.}
    \textbf{A,B}, Large-scale scanning nitrogen-vacancy (NV) microscopy of BiFeO$_3$ after poling into $+P_{[100]}$ state (\textbf{A}), and $-P_{[100]}$ state (\textbf{B}), reprinted from the previous report \cite{meisenheimer_switching_2024}.
    \textbf{C}, Magnon transport hysteresis of BiFeO$_3$ shows an amplitude $\Delta V_{ISHE} \approx 150$ nV, reprinted from the previous report \cite{harris_symmetry_2025}.
    \textbf{D,E}, Large-scale NV microscopy of 15\%-Lanthanum substituted BiFeO$_3$ (La-BFO) after poling into $+P_{[100]}$ state (\textbf{D}), and $-P_{[100]}$ state (\textbf{E}), reprinted from the previous report \cite{husain_nonvolatile_2024}. 
    \textbf{F}, Magnon transport hysteresis of La-BFO shows an amplitude $\Delta V_{ISHE} \approx 500$ nV, comparable with the current study of sBFO.
	\label{fig:figS5}
    }
\end{figure}

\clearpage
\begin{figure}
	\centering
	\includegraphics[width=\textwidth]{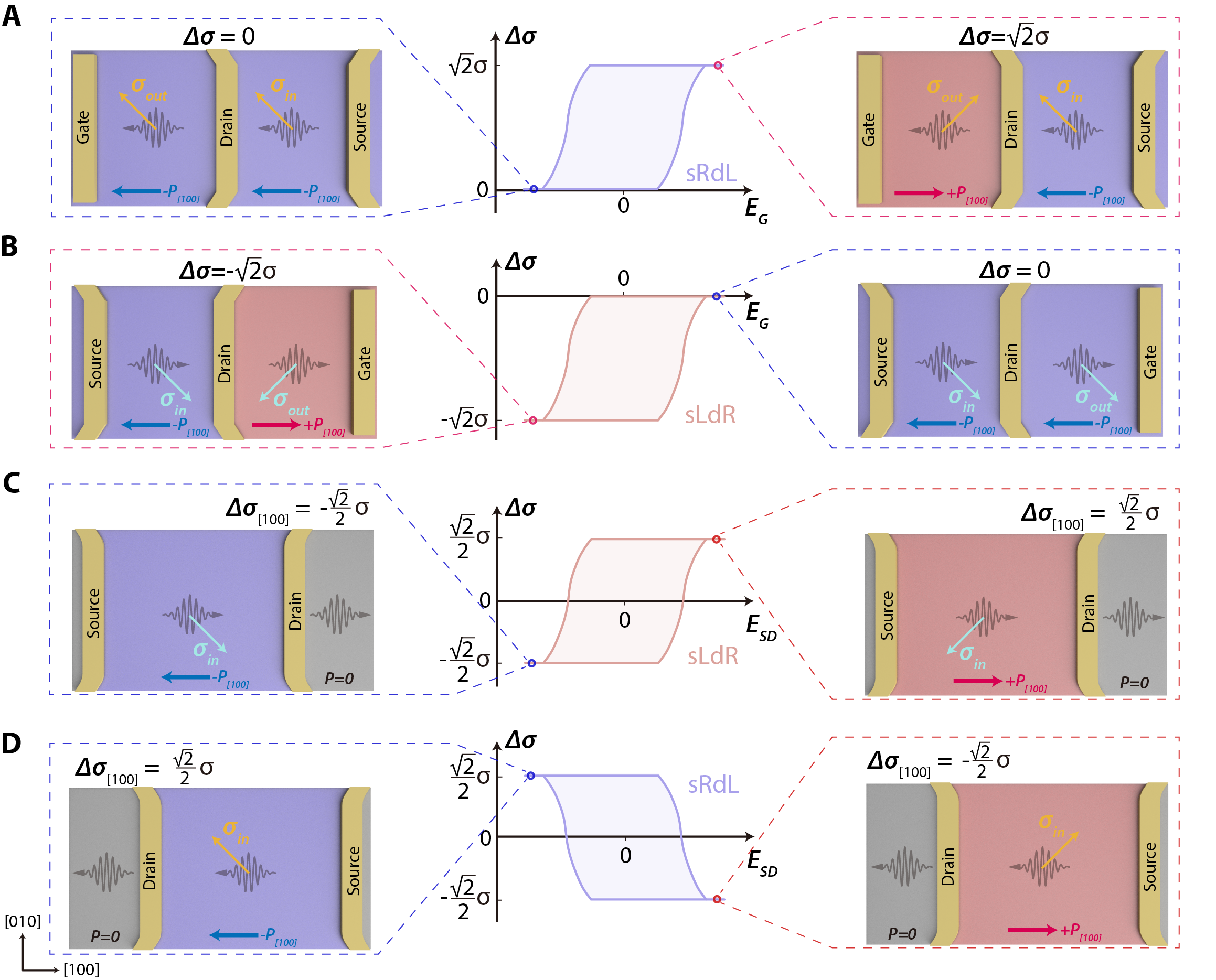} 
	\caption{\textbf{Schematic model of magnon spin angular momentum transfer at antiferromagnetic domain boundaries.}
    \textbf{A,B} Gate-field-dependent magnon spin angular momentum transfer for the (\textbf{A}) sRdL and (\textbf{B}) sLdR configurations. Uniform ferroelectric polarization produces no change in magnon spin polarization $\Delta\boldsymbol{\sigma}=0$, whereas a gate-induced antiferromagnetic domain boundary can rotate the transmitted magnon spin polarization, yielding $\Delta\boldsymbol{\sigma}=\sqrt{2}\sigma$ for the illustrated orthogonal spin orientations. Reversing the magnon propagation direction reverses the sign of the corresponding angular momentum transfer. 
    \textbf{C,D} Source–drain-field-dependent magnon spin angular momentum transfer for the (\textbf{C}) sLdR and (\textbf{D}) sRdL configurations. The region outside source-drain is unpolarized with net $P=0$. The projected angular momentum change along [100] ($\Delta\sigma_{[100]}$), accounts for the observed ISHE. The hysteresis reverses upon interchanging the source and drain, attributing to the odd-parity magnon spin splitting in cycloidal magnetic structure.
	\label{fig:figS6}
    }
\end{figure}

\clearpage
\begin{figure}
    \centering
    \includegraphics[width=1.0\textwidth]{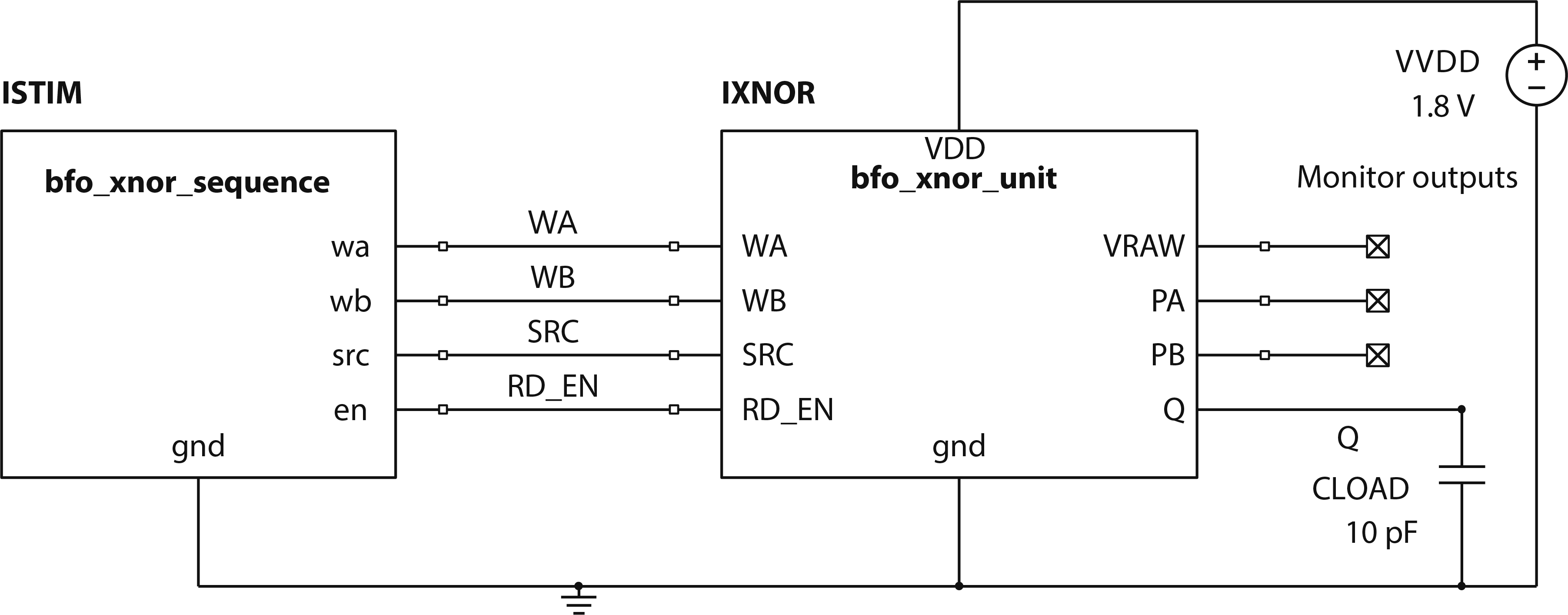}
    \caption{\textbf{Cadence testbench.} 
    ISTIM supplies the write, source and enable waveforms to IXNOR. A separate 1.8 V source powers the receiver and Q drives a 10 pF load. Crossed-square markers indicate externally unloaded monitor branches; they do not connect VRAW, PA or PB to ground. VRAW remains connected to the receiver inside IXNOR.
    \label{fig:figS7}
    }
\end{figure}

\clearpage
\begin{figure}
    \centering
    \includegraphics[width=\textwidth]{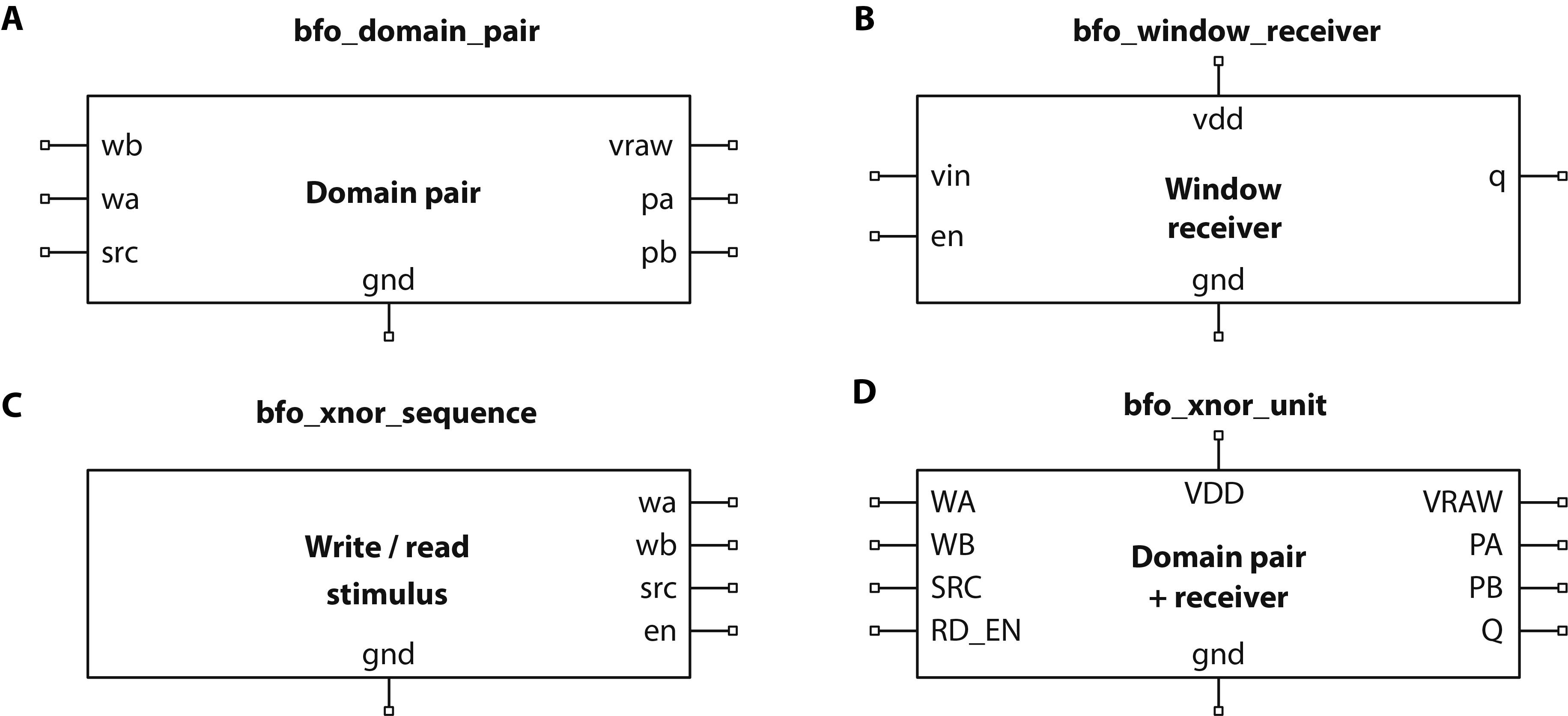}
    \caption{\textbf{Cadence interfaces.}
    \textbf{A}, Retained-state domain-pair model. \textbf{B}, Powered window receiver. \textbf{C}, Write/read stimulus generator. \textbf{D}, Composite unit containing the detector and receiver. Labels, capitalization and pin order follow the source symbols. PA and PB are diagnostic outputs; neither is connected to the receiver decision.
    \label{fig:figS8}
    }
\end{figure}

\clearpage
\begin{figure}
    \centering
    \includegraphics[width=0.8\textwidth]{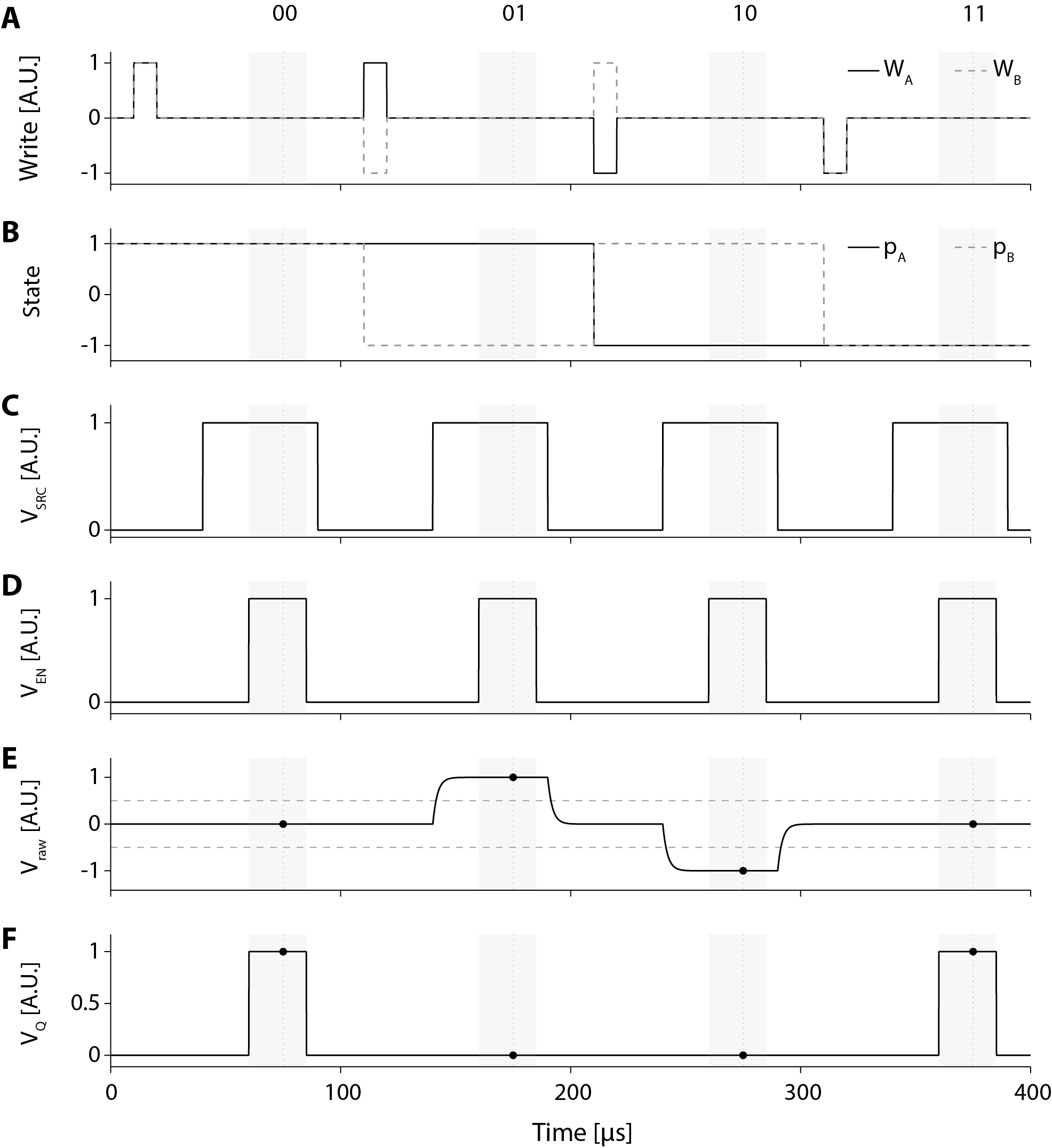}
    \caption{\textbf{Simulated transient response.}
    \textbf{A}, Write voltages (A solid, B dashed). \textbf{B}, Retained normalized states (A solid, B dashed). \textbf{C}, Read-source voltage. \textbf{D}, Read enable. \textbf{E}, Raw detector voltage, with the fixed $\pm0.5$ window marked by dashed lines. \textbf{F}, Powered receiver output. Gray bands show enabled intervals and dotted vertical lines mark sampling. Slot labels denote the states programmed after each write event. Curves connect exported samples; no smoothing is applied.
    \label{fig:figS9}
    }
\end{figure}



\clearpage
\small
\par\medskip
\begin{table}
\centering
\begin{tabular}{@{}p{0.28\linewidth}p{0.20\linewidth}p{0.44\linewidth}@{}}
\toprule Parameter & Value & Role\\\midrule
\texttt{vc\_a}, \texttt{vc\_b}&20 V each&Positive magnitudes of the write thresholds.\\
\texttt{bit\_a\_init}, \texttt{bit\_b\_init}&0, 0&Default initial logical states.\\
\texttt{tr\_state}&100 ns&State-monitor transition smoothing.\\
\texttt{r\_write}&1 T$\Omega$&Leakage resistance per write terminal.\\
\texttt{c\_write}&1 fF&Capacitance per write terminal.\\
\texttt{r\_src}&1 k$\Omega$&Source/heater electrical resistance.\\
\texttt{i\_ref}&3 mA&Read-current normalization.\\
\texttt{tau\_read}&2 $\mu$s&Assumed read-envelope time constant.\\
\texttt{v\_contrast}&0.6 $\mu$V&Nominal unequal-state contrast at unit drive.\\
\texttt{v\_common}&0 V&Common-state response coefficient.\\
\texttt{v\_offset}&0 V&Additive detector offset.\\
\texttt{r\_det}&1 k$\Omega$&Detector source resistance.\\
\texttt{polarity}&$+1$&Fixed signed measurement convention.\\\bottomrule
\end{tabular}
\caption{\textbf{Defaults in} \texttt{bfo\_domain\_pair.va}.}
\label{tab:tabS1}
\end{table}
\par\medskip

\clearpage
\par\medskip
\begin{table}
\centering
\begin{tabular}{@{}p{0.28\linewidth}p{0.20\linewidth}p{0.44\linewidth}@{}}
\toprule Parameter & Value & Role\\\midrule
\multicolumn{3}{l}{\textit{Window receiver}}\\
\texttt{v\_center}&0 V&Window center.\\
\texttt{v\_window}&0.3 $\mu$V&Window half-width.\\
\texttt{v\_enable}&0.9 V&Read-enable threshold.\\
\texttt{r\_in}&1 T$\Omega$&Input and enable resistive loading.\\
\texttt{c\_in}&1 pF&Raw-input capacitance.\\
\texttt{r\_on}, \texttt{r\_off}&100 $\Omega$, 1 T$\Omega$&Output conductance limits.\\
\texttt{t\_edge}&100 ns&Decision-to-conductance transition smoothing.\\\midrule
\multicolumn{3}{l}{\textit{Stimulus generator}}\\
\texttt{slot\_time}&100 $\mu$s&Duration of each programmed case.\\
\texttt{v\_write}&40 V&Write-pulse magnitude; polarity encodes the bit.\\
\texttt{v\_source}&3 V&Read-source voltage.\\
\texttt{v\_enable\_high}&1.8 V&High enable level.\\
\texttt{t\_edge}&100 ns&Stimulus transition smoothing.\\\midrule
\multicolumn{3}{l}{\textit{External testbench}}\\
VDD&1.8 V&Reader output-stage supply.\\
CLOAD&10 pF&Load on Q.\\\bottomrule
\end{tabular}
\caption{\textbf{Receiver, stimulus and external-load defaults.}}
\label{tab:tabS2}
\end{table}
\par\medskip
\normalsize

\clearpage
\begin{table}
\centering\small
\begin{tabular}{@{}ccccrrrr@{}}\toprule
$AB$ & $t$ ($\mu$s) & $p_A$ & $p_B$ & $V_{\mathrm{raw}}$ ($\mu$V) & $V_Q$ (V) & Q bit & XNOR\\\midrule
00&75 &+1&+1& 0.000&1.800&1&1\\
01&175&+1&-1&+0.600&0.000&0&0\\
10&275&-1&+1&-0.600&0.000&0&0\\
11&375&-1&-1& 0.000&1.800&1&1\\\bottomrule
\end{tabular}
\caption{\textbf{Samples from the exported transient.} Q is decoded at 0.9 V while read enable is high. Displayed voltages are rounded.}
\label{tab:tabS3}
\end{table}

\end{document}